\documentclass[aps,prb,twocolumn,showpacs]{revtex4-1}
\usepackage[utf8]{inputenc}
\usepackage{amssymb}
\usepackage{amsmath}
\usepackage{amsfonts}
\usepackage{dsfont}
\usepackage[usenames,dvipsnames]{xcolor}
\usepackage[pdftex]{graphicx}
\usepackage[pdftex,plainpages=false,colorlinks=true,linkcolor=Red, citecolor=blue, urlcolor=blue]{hyperref}
\usepackage{bm}

\renewcommand{\vec}[1]{\bm{\mathrm{#1}}}

\begin{document}

\title{Density Waves Cause Sub-Gap Structures\\ but no Pseudogap in Superconducting Cuprates }
\author{S. \surname{Verret}}
\affiliation{D\'epartement de physique, Universit\'e de Sherbrooke, Qu\'ebec, Canada  J1K 2R1}
\author{M. \surname{Charlebois}}
\affiliation{D\'epartement de physique, Universit\'e de Sherbrooke, Qu\'ebec, Canada  J1K 2R1}
\author{D. \surname{S\'en\'echal}}
\affiliation{D\'epartement de physique, Universit\'e de Sherbrooke, Qu\'ebec, Canada  J1K 2R1}
\author{A.-M. S. \surname{Tremblay}}
\email[Corresponding author: ]{andre-marie.tremblay@usherbrooke.ca}
\affiliation{D\'epartement de physique, Universit\'e de Sherbrooke, Qu\'ebec, Canada  J1K 2R1}
\date{\today}
\begin{abstract}
In scanning tunneling microscopy (STM) conductance curves, the superconducting gap of cuprates is sometimes accompanied by small sub-gap structures at very low energy. This was documented early on near vortex cores and later at zero magnetic field. Using mean-field toy models of coexisting $d$-wave superconductivity ($d$SC), \emph{d}-form factor density wave ($d$FF-DW), and extended s-wave pair density wave ($s'$PDW), we find agreement with this phenomenon, with $s'$PDW playing a critical role. We explore the high variability of the gap structure with changes in band structure and density wave (DW) wave vector, thus explaining why sub-gap structures may not be a universal feature in cuprates. In the absence of nesting, non-superconducting results never show signs of pseudogap, even for large density waves magnitudes, therefore reinforcing the idea of a distinct origin for the pseudogap, beyond mean-field theory. Therefore, we also briefly consider the effect of DWs on a pre-existing pseudogap. 
\end{abstract}

\maketitle

\section{Introduction}

The presence of density waves (DWs) in high temperature superconducting cuprates is now well-established~\cite{ghiringhelli_long-range_2012, comin_resonant_2016}, and growing evidence suggest that DWs and the pseudogap (PG) are related~\cite{comin_charge_2014} but distinct phenomena~\cite{alloul_what_2014, atkinson_charge_2015, badoux_change_2016}. However, it is not yet clear how this distinction between DWs and the PG appears in the tunneling density of state (DOS) of cuprates~\cite{he_fermi_2014, hamidian_atomic-scale_2016}.

Scanning tunneling microscopy played a key role in the discovery of DWs in cuprates. The early finding of a checkerboard DW in vortex cores~\cite{hoffman_four_2002}, where $d$-wave superconductivity ($d$SC) is weakened by a magnetic field, suggested a competition between $d$SC and DWs. However, finding DWs was the culmination of much work previously focusing on the presence of low-energy \emph{sub-gap structures} in the local DOS surrounding vortex cores~\cite{fischer_scanning_2007}. Sub-gap structures were typically found between $\pm$5~meV and $\pm$10~meV in the conductance spectra of optimally doped YbBa$_{2}$Cu$_{3}$O$_{7-\delta}$ (YBCO)~\cite{maggio-aprile_direct_1995} and Bi$_2$Sr$_2$CaCu$_2$O$_8$~(BSCCO)~\cite{pan_stm_2000, hoogenboom_low-energy_2000}.

\begin{figure}
\includegraphics{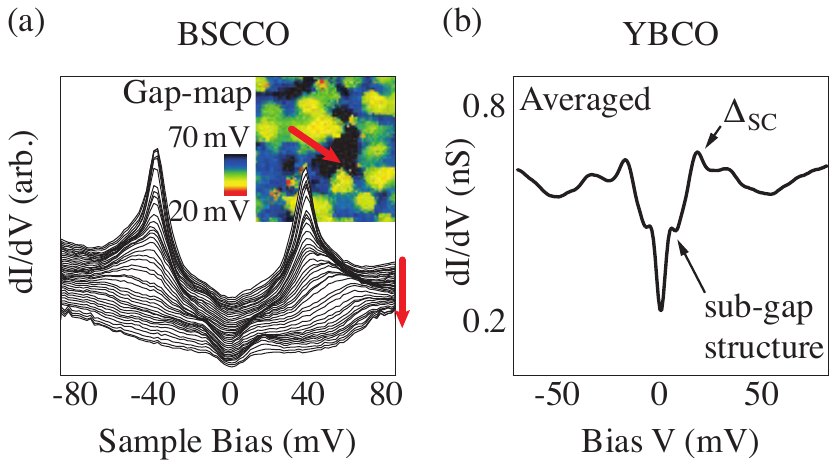}
\caption{Sub-gap structures seen at zero magnetic field, {\bf a)} in the inhomogeneous conductance spectra in BSCCO (adapted from Ref.~\onlinecite{mcelroy_homogenous_2004}, preprint version of Ref.~\onlinecite{mcelroy_coincidence_2005})
{\bf b)} and in the averaged spectra for~YBCO~(adapted from~\cite{bruer_revisiting_2016})}
\label{bruer_measure}
\end{figure}

The occurrence of sub-gap structures (SGS) exactly where charge order was found indicates a likely relation between these phenomena~\cite{levy_fourfold_2005}. Moreover, sub-gap structures are also found at zero-field, so vortices and magnetic fields are not the key to explain them. Bru\`er \emph{et al.} recently reported those structures in the averaged zero-field spectra of YBCO, and suggested: ``\emph{it is tempting to link the SGS with the static charge density wave discovered recently in Y123}''~\cite{bruer_revisiting_2016}. Equivalent zero-field structures had been extensively studied in underdoped samples of BSCCO~\cite{mcelroy_coincidence_2005, kohsaka_intrinsic_2007, hamidian_atomic-scale_2016} and mainly occur in locally resolved spectra where density waves are enhanced. Fig.~\ref{bruer_measure} reproduces typical examples of sub-gap structures at zero-field in BSCCO and YBCO.

Along with proposed scenarios of spin density waves~\cite{andersen_checkerboard_2003} and staggered flux~\cite{kishine_staggered_2001, lee_vortex_2001}, early theoretical work firmly stated the likeliness of pair density waves (PDWs)~\cite{chen_antiferromagnetism_2002, chen_pair_2004, seo_$d$-wave_2007, seo_complementary_2008} to explain the checkerboard patterns found in STM. Recent work by Agterberg and Garaud further showed that for competing $d$SC and PDW, a vortex core will favor PDW through the suppression of $d$SC, and also drive complementary charge modulation, as those seen in experiments~\cite{agterberg_checkerboard_2015}. After the observation by Lee that some photoemission results in cuprates agree better with a PDW scenario than with one based on charge density waves~\cite{lee_amperean_2014}, and after much work stating how many theoretical models are prone to several forms of PDWs~\cite{himeda_stripe_2002, raczkowski_unidirectional_2007, wang_coexistence_2015, freire_renormalization_2015, fradkin_colloquium_2015}, experimental evidence of pair density wave was finally obtained, at zero-field, for an underdoped sample of BSCCO, through scanning Josephson tunneling spectroscopy~\cite{hamidian_detection_2016}. It would be hard, at this point, to exclude PDWs from the cuprate puzzle.

In this work, we consider phenomenological mean-field Hamiltonians for coexisting bond-centered density wave and pair density waves (similar to those experimentally reported in Ref.~\onlinecite{fujita_direct_2014,hamidian_detection_2016}) combined to d-wave superconductivity, and we find qualitative agreement with observed spectra for sub-gap structures in the~DOS.

Our main conclusions are that (i) pair density waves are a key ingredient to obtain low-energy sub-gap structures in the $d$SC gap, and (ii) no pseudogap appears in the DOS for systems with charge or pair density waves alone and without nesting, clarifying the need for a distinct phenomenon to explain the tunneling PG. These conclusions hold for a range of established band structures for cuprates, and realistic magnitudes for the $d$SC and DWs mean fields, as well as for unidirectional and bidirectional DWs.

A secondary conclusion of our work is that the band structure plays a role at least as critical as the periodicity of the DW in the reconstruction caused by DW, and in the corresponding~DOS. The calculated DOS exhibit extraordinary complexity and variability that reflect the great variability observed in conductance spectra in STM~\cite{mcelroy_coincidence_2005, kohsaka_intrinsic_2007, machida_bipartite_2016}, probably explaining why sub-gap structures are not always seen. This effect does not seem to be discussed in previous studies~\cite{allais_connecting_2014, seo_complementary_2008}, and needs to be considered carefully when attempting to find universal properties among different compounds with band reconstruction.

The model, including the description of the various DWs, is described in Sec.~\ref{Sec:Model}. After a brief discussion of the method in Sec.~\ref{Sec:Method}, we describe the results in Sec.~\ref{Sec:Results} with additional discussion in Sec.~\ref{Sec:Discussion}, including, in Sec.~\ref{Sec:WithPseudogap}, a few additional calculations and discussion of how DWs sub-gap structures may modify the DOS in cases where a pseudogap~\cite{yang_phenomenological_2006,qi_effective_2010} is pre-existing. 

\section{Model}\label{Sec:Model}

Even assuming that sharp quasiparticles reappear in ordered states~\cite{vishik_arpes_2010}, allowing for a single-particle approach, modeling density waves in cuprates is a difficult theoretical problem for two reasons: (i) they are incommensurate, and (ii) they are short-ranged, meaning that the system is disordered. Numerous recent work address these issues (see~Ref.~\onlinecite{sachdev_bond_2013, allais_connecting_2014, lee_cold-spots_2016} and references therein).

By contrast, the computations presented here are for simple, long-ranged commensurate order. 
In the past, similar simplifications were used to successfully obtain DW-induced quantum oscillation frequencies in YBCO~\cite{allais_connecting_2014},
and also to address the closely related question of how spin density waves can reconcile the photoemission and tunneling observations in cuprates~\cite{andersen_extinction_2009}.
The interest of such simplifications comes from the wide range of parameters that can be tested. Our results, along with those of Ref.~\onlinecite{allais_connecting_2014}~and~\onlinecite{andersen_extinction_2009}, suggest that the realistic disordered incommensurate systems may be approximated locally by ordered commensurate ones. The complex variation of the local DOS with position in experiments~\cite{kohsaka_intrinsic_2007} would be reflected, in our model, by the complex variation of the DOS with band structure, DWs wave vector, and the strengths of DWs mean fields.
It is therefore an approach in the spirit of the virtual crystal approximation~\cite{nordheim_electron_1931, bellaiche_virtual_2000}, which is computationally much cheaper than solving a large disordered system. 
Notably, in Ref.~\onlinecite{andersen_extinction_2009}, it was shown, along with results comparable to ours, but for spin DW, that adding disorder to long-ranged commensurate DW systems simply spreads otherwise sharp features found in the spectra, without changing the qualitative conclusions. This approach finds additional justification from work showing how strong correlations generically protect $d$wave superconductivity against disorder~\cite{garg_strong_2008, tang_strong_2016}.

The model we study is a 2D tight-binding mean-field one-band Hamiltonian at zero temperature on a square lattice with four terms:
\begin{align}
H=H_{0}+H_{\text{$d$SC}}+H_{\text{$d$FF-DW}}+H_{\text{$s'$PDW}}.
\label{real_space_hamiltonian}
\end{align}

\begin{figure}
\centering
\includegraphics{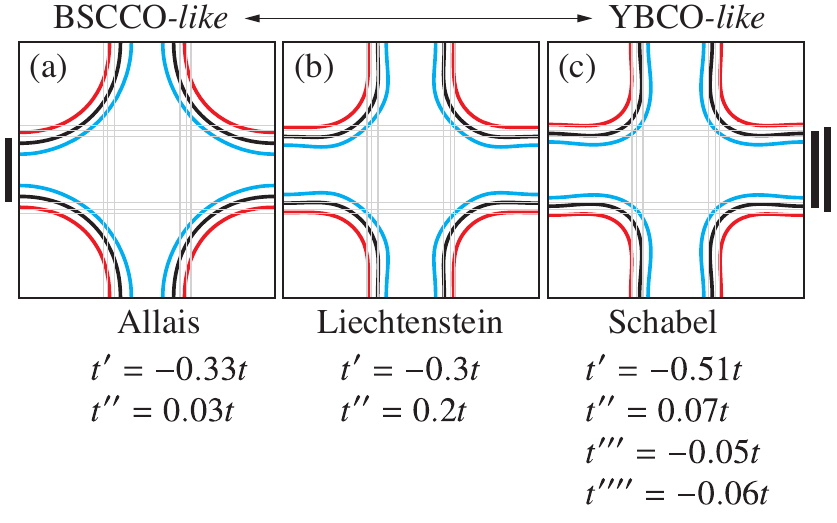} %[width=8.5cm]
\caption{Three sets of band parameters and their Fermi surfaces at half-filling (red curves), doping $p=0.125$ (black curves) and doping $p=0.25$ (blue curves). Parameters in {\bf a)}, from Allais~\emph{et al.}~\cite{allais_connecting_2014}, were used to describe quantum oscillations frequencies observed in YBCO when reconstructed by dFF-DW. However, the Fermi-surface resembles that obtained for BSCCO from photoemission experiments~\cite{norman_phenomenological_1995}. Parameters in \textbf{b)}, first used by Liechtenstein~\cite{liechtenstein_quasiparticle_1996}, but also in Yang-Rice-Zhang theory, and usually preferred~\cite{kancharla_anomalous_2008, yang_phenomenological_2006}, are based on \emph{ab initio} calculations to represent a fairly generic one-band cuprate~\cite{mattheiss_electronic_1990, andersen_lda_1995, pavarini_band-structure_2001}. Parameters in \textbf{c)}, from Schabel~\emph{et al.}~\cite{schabel_angle-resolved_1998}, more rarely used, come from photoemission data on YBCO. It is important to recall that both BSCCO and YBCO are bi-layer compounds, so that any one-band model is a rough approximation~\cite{markiewicz_one-band_2005, hoogenboom_modeling_2003}. Thin gray lines in plots correspond to reduced Brillouin-zones while bold lines on the sides correspond to the length of the DWs wave vectors; from left to right,~$\frac{q}{L}=\frac{1}{4},\frac{3}{10},\frac{1}{3}$.}
\label{band_structures}
\end{figure}

The first term is the underlying band structure:
\begin{align}
H_0=&
\sum_{\vec r,\vec a,\sigma}
t_{\vec a}
c^\dagger_{\vec r+\vec a,\sigma}c_{\vec r,\sigma}
- \mu\sum_{\vec r,\sigma} 
c^\dagger_{\vec r,\sigma}c_{\vec r,\sigma}.
\end{align}
Operators $c^{\dagger}_{\vec r, \sigma}$ and $c_{\vec r, \sigma}$ respectively create and annihilate electrons at position $\vec r$ with spin $\sigma$. The sum on $\vec{a}$ spans all lattice vectors (pointing to all neighbors of $\vec r$), and the parameters $t_{\vec a}$ are the real-space components of the tight-binding dispersion $\xi(\vec k) = \epsilon(\vec k)-\mu$. Those are typically denoted $t$ for first neighbor hopping, $t'$ for second neighbors, etc. \emph{Ab initio} calculations~\cite{andersen_lda_1995, liechtenstein_quasiparticle_1996, pavarini_band-structure_2001, markiewicz_one-band_2005} and photoemission experiments~\cite{norman_phenomenological_1995, markiewicz_one-band_2005, he_single-band_2011} prescribe an acceptable range of band-structures for cuprates, from which we chose three representative sets, depicted in detail in Fig.~\ref{band_structures}. 
For easy comparison with experiments, we express $\omega$ in electron volts~(eV), using the energy-scale~$t=250$~meV. The chemical potential $\mu$ is always set so that the density yields $p=0.125$ of hole doping relative to a half-filled band. 

The second term in the Hamiltonian represents $d$-wave superconductivity~($d$SC):
\begin{align}
H_{\text{$d$SC}}=&
\sum_{\vec r,\vec a}
\tfrac{1}{2}\Delta_{\vec a}
\big(c_{\vec r+\vec a,\uparrow}c_{\vec r,\downarrow}-c_{\vec r+\vec a,\downarrow}c_{\vec r,\uparrow}\big)
+\text{H.c.}
\end{align}
The $d$-wave gap $\tfrac{\Delta}{2}(\cos k_x-\cos k_y)$ is obtained by setting $\Delta_{\hat{\vec x}}=-\Delta_{\hat{\vec y}}\equiv\Delta/2$, resulting in coherence peaks located at~$\pm\Delta$. 
%With $t=250$~meV, 
We adjust the mean-field $\Delta$ phenomenologically; the gap magnitude that fits experimental tunneling spectra~\cite{fischer_scanning_2007} is around \mbox{$\Delta=0.12t=30$~meV}, close to that of YBCO at optimal doping.

\begin{figure}
\centering
\includegraphics{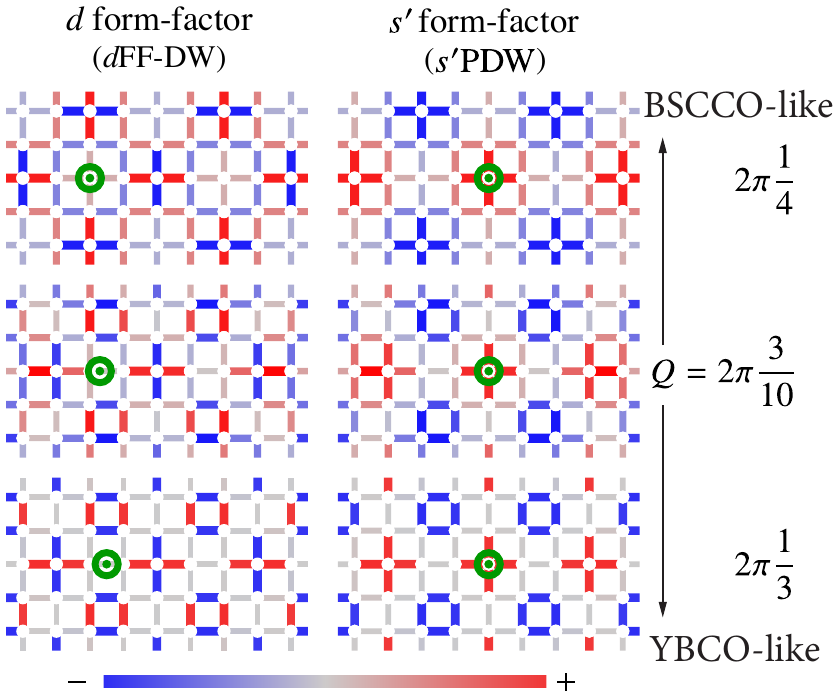} %[width=8.5cm]
\caption{Illustration of the form factors for each bidirectional bond-centered density waves used, with wave vectors~$\frac{q}{L}=\frac{1}{4},\frac{3}{10},\frac{1}{3}$. Each tilling illustrate the quantity $\text{e}^{\text{i}\vec Q(\vec r+\frac{\vec a}{2})}$ which modulates nearest-neighbour hoping in Eq.~\ref{eqdffdw}, for the case of $d$FF-DW, and nearest-neighbour pairing in in Eq.~\ref{eqspdw}, for the case of $s'$PDW. On the left side, the $d$FF-DW form factor, $t_{\vec Q,\vec a}$, defined with $t_{\vec Q,\pm\hat{\vec{x}}}=-t_{\vec Q,\pm\hat{\vec{y}}}$, modulates the hopping with opposite signs on the $x$ and $y$ bonds. By contrast, on the right side, the $s'$PDW form factor, $\Delta_{\vec Q,\vec a}$, defined with $\Delta_{\pm\vec Q,\pm\hat{\vec{x}}}=\Delta_{\pm\vec Q,\pm\hat{\vec{y}}}$ modulates the pairing with same sign in both directions. 
These modulated hopping (pairing) terms can be seen as the corresponding effective static electric (pairing) potential applied on top of the original crystal potential. Although both the lattice potential and the DW potentials considered are C$_4$ symmetric, two overlapping periodic C$_4$ potentials only preserve C$_4$ symmetry when their symmetry centers coincide. As one can see here, symmetry centers (green circles) of the form factors do not always coincide with those of the lattice (which could be either the lattice sites or the corners of the Wigner-Seitz cells), and thus C$_4$ symmetry is broken for $\frac{q}{L}=\frac{3}{10}$ (middle left) and $\frac{1}{3}$ (bottom left). For $\frac{q}{L}=\frac{1}{4}$ (top left), symmetry is preserved since the symmetry center falls back on a lattice site. In the case of $s'$PDW, C$_4$ symmetry is always preserved. Let us also specify that shifting the DW potential by a carefully chosen phase would preserve C$_4$ symmetry in all cases.}
\label{DWpatterns}
\end{figure}

The third term represents bond-centered density waves:
\begin{align}
H_{\text{$d$FF-DW}}=&
\sum_{\vec r,\vec a,\sigma}
\sum_{\vec Q}
\tfrac{1}{2}t_{\vec Q,\vec a}
\text{e}^{\text{i}\vec Q(\vec r+\frac{\vec a}{2})}
c^\dagger_{\vec r+\vec a,\sigma}c_{\vec r,\sigma}
+\text{H.c.}
\label{eqdffdw}
\end{align}
We consider a bidirectional $d$-form-factor density wave~($d$FF-DW) following Refs.~\onlinecite{fujita_direct_2014,allais_connecting_2014} by defining $t_{\vec Q,\pm\hat{\vec{x}}}=-t_{\vec Q,\pm\hat{\vec{y}}}\equiv t_Q/2$ (see Fig.~\ref{DWpatterns}). The Hermitian conjugate naturally includes density waves of opposite wave vectors.

The fourth and last term represents pair density waves:
\begin{align}
H_{\text{$s'$PDW}}=&
\sum_{\vec r,\vec a,\vec Q}
\tfrac{1}{2}\Delta_{\vec Q,\vec a}
\text{e}^{\text{i}\vec Q(\vec r+\frac{\vec a}{2})}
\big(c_{\vec r+\vec a,\uparrow}c_{\vec r,\downarrow}-c_{\vec r+\vec a,\downarrow}c_{\vec r,\uparrow}\big)
+\text{H.c.}
\label{eqspdw}
\end{align}
Following experimental evidence, we consider \mbox{extended-$s$} singlet pair density wave ($s'$PDW)~\cite{hamidian_detection_2016}, defined by $\Delta_{\pm\vec Q,\pm\hat{\vec{x}}}=\Delta_{\pm\vec Q,\pm\hat{\vec{y}}} \equiv \Delta_Q/2$ (see Fig.~\ref{DWpatterns}). In this case, opposite wave vectors correspond to pairs carrying opposite momentum and must be included separately.

Both $d$FF-DW and $s'$PDW are modulated according to wave vectors $\vec Q_x=\tfrac{q}{L}2\pi\hat{\vec x}$ and/or $\vec Q_y=\tfrac{q}{L}2\pi\hat{\vec y}$. All results \emph{shown} in this paper are for the purely bidirectional case, \emph{i.e.} with $t_{\vec Q_x,{\vec{a}}}=t_{\vec Q_y,{\vec{a}}}$ (illustrated in Fig.~\ref{DWpatterns}), but we verified that all our conclusions also hold in the unidirectional case, \emph{i.e.} with $t_{\vec Q_x,{\vec{a}}}\neq 0,  t_{\vec Q_y,{\vec{a}}}=0$.
In real samples, the bidirectional or unidirectional character of the DWs is not all black or white.
Although both directions of oscillations are found, analysis of resonant x-ray scattering in YBCO proved to be more consistent with domains of unidirectional character~\cite{comin_broken_2015}. Comparable conclusions were obtained from the analysis of STM results in BSCCO~\cite{hamidian_atomic-scale_2016}, with the additional nuance that those predominantly unidirectional domains tend to overlap at the nanoscale level, explaining earlier reports of checkerboard patterns. As we will repeat throughout the paper, the set of results obtained here for bidirectional DWs is undistinguishable from that of purely unidirectional DWs; the main clear difference is that recognizable structures in the density of states are always stronger in the bidirectional case, resembling a lot what one would expect from simply adding up the effect of waves in both directions.

For the period of the DW, we consider rational fractions $\tfrac{q}{L}$ that correspond to $q$ periods of the density wave over $L$ unit cells.  More specifically, we choose $\tfrac{1}{4}$, $\tfrac{1}{3}$, and $\tfrac{3}{10}$ for $\tfrac{q}{L}$ that are the same for charge and pair density waves. For doping $p=0.125$, the values $\tfrac{q}{L}=\tfrac{1}{4}$ and $\tfrac{1}{3}$ are close to values measured in BSCCO and YBCO, respectively~\cite{comin_resonant_2016}, and $\tfrac{q}{L}=\tfrac{3}{10}$ is a manageable fraction in between, to test low commensurability. Those wave vectors are shown as bold lines, left and right of the Fermi surface plots in~Fig.~\ref{band_structures}, with the corresponding reduced Brillouin zone boundaries shown as pale gray lines. Experimentally, wave-vectors do not perfectly nest flat parts of the Fermi surface, so we do not consider this case~\cite{comin_charge_2014,mesaros_commensurate_2016}~\footnote{Non-nesting DW scenario can be justified given a sufficiently complicated interaction function $U(\vec q)$, because then the largest susceptibility $\chi_0 / (1-U(\vec q)\chi_0(\vec q))$ is not necessarily at a nesting wave vector of the Fermi surface.}.

The magnitudes of mean fields $t_Q$~($d$FF-DW) and $\Delta_Q$~($s'$PDW) are harder to determine than $\Delta$~($d$SC). In their study of quantum oscillations stemming from the Fermi-surface reconstruction by charge density waves, Allais \emph{et al.}~\cite{allais_connecting_2014} fixed a $d$FF-DW with $t_Q$ equivalent to $0.3t$ in our notation. That was to represent the enhanced DW under high magnetic field. Here, we consider values ranging from $t_Q=0.04t$ to $0.28t$. For $s'$PDW, the Josephson tunneling critical current was reported to oscillate in space at 5\% of its homogeneous value~\cite{hamidian_detection_2016}. Here we consider values between $\Delta_Q=0.04t$ and $0.28t$ that would mimic strong inhomogeneity.

\section{Method}\label{Sec:Method}

For commensurate wave vectors of denominator $L$, the $\vec k$-space representation of Hamiltonian \eqref{real_space_hamiltonian} can be expressed as a finite matrix $\hat{\text{H}}(\vec k)$ of dimension $2L^2\times 2L^2$ for bidirectional DW and  $2L\times 2L$ for unidirectional DW, with the factor 2 accounting for spin and pairing through Nambu formalism. The chosen basis is defined by the following Nambu spinor:
\begin{align}
\begin{pmatrix}
c^{\dagger}_{\vec k,\uparrow} & ... & c^{\dagger}_{\vec k+n\vec Q_x+m\vec Q_y,\uparrow} & 
... & c_{-\vec k,\downarrow} & 
... & c_{-\vec k-n\vec Q_x-m\vec Q_y,\downarrow} & ... 
\end{pmatrix},
\end{align}
$m$ and $n$ being integers between~$0$~and~$L$, so that each diagonal term of $\hat{\text{H}}(\vec k)$ corresponds to a displaced reduced Brillouin zone (rBZ).
The position averaged local DOS is then obtained from
%\begin{align}
%\text{DOS}(\omega) = -\int\limits_{\text{rBZ}}\text{d}\vec k\ \text{tr}\Bigg[ 
%\underbrace{
%\frac{1}{\pi}\text{Im}\Bigg\{\frac{1}{\omega + \text{i}\eta - \hat{\text{H}}(\vec k)} \Bigg\}
% (\tau_3 \otimes \mathds{1})
%}_{\equiv\hat{\text{A}}(\vec k,\omega)}
%\Bigg],
%\label{eq_dos}
%\end{align}
\begin{align}
\text{DOS}(\omega) &= \int\limits_{\text{rBZ}}\text{d}\vec k\ \text{tr}_{\uparrow}\big[ 
\hat{\text{A}}(\vec k,\omega)
\big],
\label{eq_dos}
\end{align}
with:
\begin{align}
\hat{\text{A}}(\vec k,\omega)&\equiv
-2\text{Im}\Bigg\{\frac{1}{\omega + \text{i}\eta - \hat{\text{H}}(\vec k)} \Bigg\}
,
\label{eq_akw}
\end{align}
where the integral extends over all $\vec k$-vectors for one rBZ and the trace sums contributions from all bands (equivalent to summing the local DOS from each Cu atoms of a $L\times L$ unit-cell). The trace sums only the spin up part of the Nambu representation, a subtlety indicated by the $\uparrow$ subscript.
%(equivalent to take the partial trace on the $L^2$ subspace and only keep the first diagonal element of the remaining $2\times2$ Nambu matrix).
The term inside the trace is the momentum-dependent spectral weight $\hat{\text{A}}(\vec k, \omega)$, which we use to show the reconstructed Fermi surface ($\omega = 0$). To do so, we rebuild the content of the original Brillouin zone using the diagonal elements of the matrix $\hat{\text{A}}(\vec k, \omega)$, again counting only spin up terms.
The only approximation is the choice of a manageable numerical $\eta$ (between 0.05 and 0.005, specified for each figure), broadening the electronic dispersion so that the Dirac function $\delta(\omega~-~E(\vec k))$ becomes a Lorentzian~$\eta/(\pi[(\omega-E(\vec k))^2-\eta^2])$.

\begin{figure*}
\includegraphics{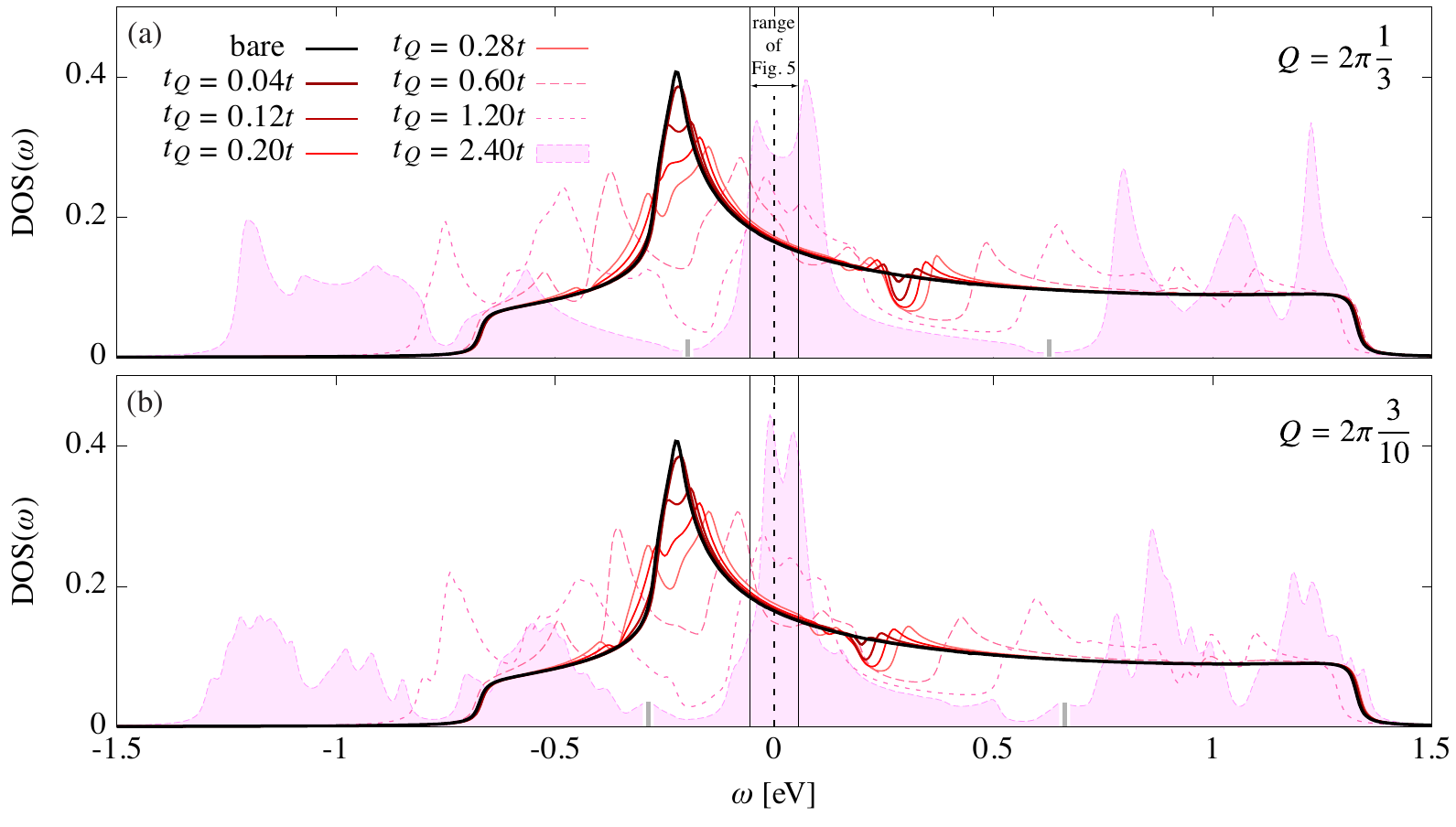} %[width=18cm]
\caption{Typical examples for the complete density of states (DOS) in the presence of bidirectional $d$-form-factor density wave (dFF-DW) with various magnitude for wave vector of (a) $Q=2\pi\frac{1}{3}$ (top panel) and (b) $Q=2\pi\frac{3}{10}$ (bottom panel) with the band of Liechtenstein (Fig.~\ref{band_structures}(b)). No superconductivity nor pair density wave is applied here. The low energy range considered for Fig.~\ref{figure_variability} (below 100 meV) is delimited by black lines and the Fermi energy (at $\omega = 0$, for doping $p=0.125$) is marked by the black dashed line. Instead of a pseudogap, nothing happens close to the Fermi energy for weak DWs, and a broad structure appears for strong DWs. The depletion of states in the middle of this structure must not be confused for a pseudogap as its dependence on $t_Q$ is inconsistent with experiments (see first paragraph of \emph{results}).
The black line corresponds to the bare DOS (no $d$FF-DW), thiner red shaded lines from black to pink correspond to the DOS in presence of $d$FF-DW with different $t_Q$ (see the legend). We used dashed lines for $t_Q=0.6t$ and higher to improve clarity.
The actual gaps from $d$FF-DW appears progressively at $\sim\pm250$ meV (corresponding to $\pm t$), irrespective of the value of $t_Q$. Small gray marks separate the shaded pink DOS, for $t_Q=2.4t$, in three regions each containing a $1/3$ of the states. This shows that the gaps of bidirectional $d$FF-DWs appear close to fillings $1/3$ and $2/3$ on the band. In the case of wave vector $Q=2\pi\frac{1}{3}$, the gaps are \emph{exactly} at fillings $1/3$ and $2/3$ (see discussion). The similarity between panels (a) and (b) indicates that wave vectors $1/3$ and $3/10$ yield similar physical consequences despite different commensurability. For all curves, Lorentzian broadening is $\eta=0.05t$, with $t=250$~meV.}
\label{figure_full_dos}
\end{figure*}

\begin{figure*}
\includegraphics{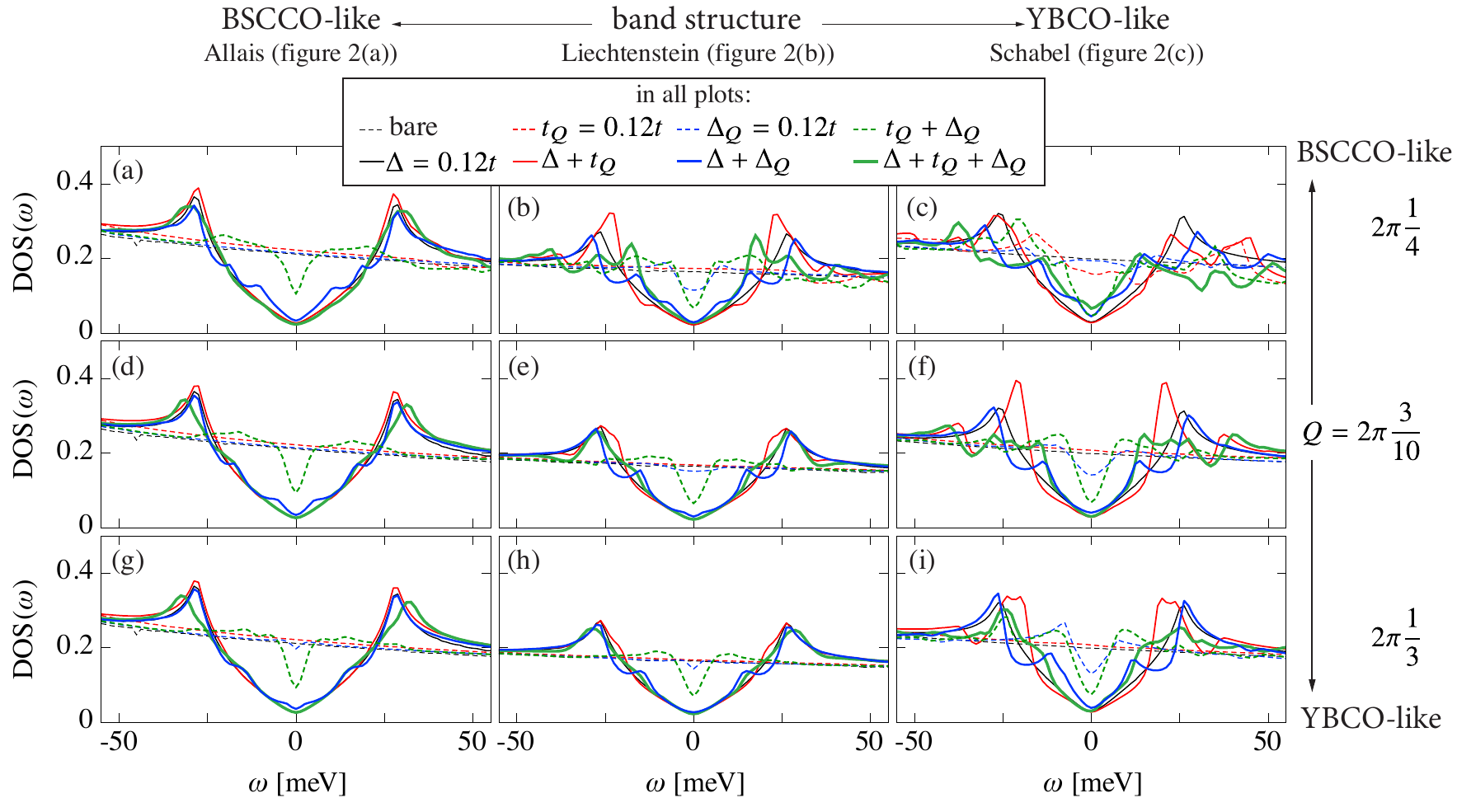} %[width=18cm]
\caption{Great variability in density of states (DOS) gapped by $d$-wave superconductivity ($d$SC, with $\Delta=0.12t$), and reconstructed by bond-centered $d$-form factor density wave ($d$FF-DW with $t_Q=0.12t$) and/or $s'$-form factor density wave ($s'$PDW with $\Delta_Q=0.12t$). Rows display the DOS for fixed DW wave vector and different band structures; columns display DOS with fixed band structure and different DW wave vectors. In each plot, the dotted lines are without superconductivity, for the bare band (dotted black), $d$FF-DW alone (dotted red), $s'$PDW alone (dotted blue), and $d$FF-DW coexisting with $s'$PDW (dotted green). Solid lines all include $d$-wave superconductivity; bare band alone with $d$SC (thinnest solid black), $d$SC with $d$FF-DW only (thin solid red), $d$SC with $s'$PDW only (thicker solid blue), and $d$SC with both $d$FF-DW and $s'$PDW (thickest solid green). For all curves, Lorentzian broadening is $\eta=0.005t$ with $t=250$~meV.}
\label{figure_variability}
\end{figure*}

\begin{figure*}
\includegraphics{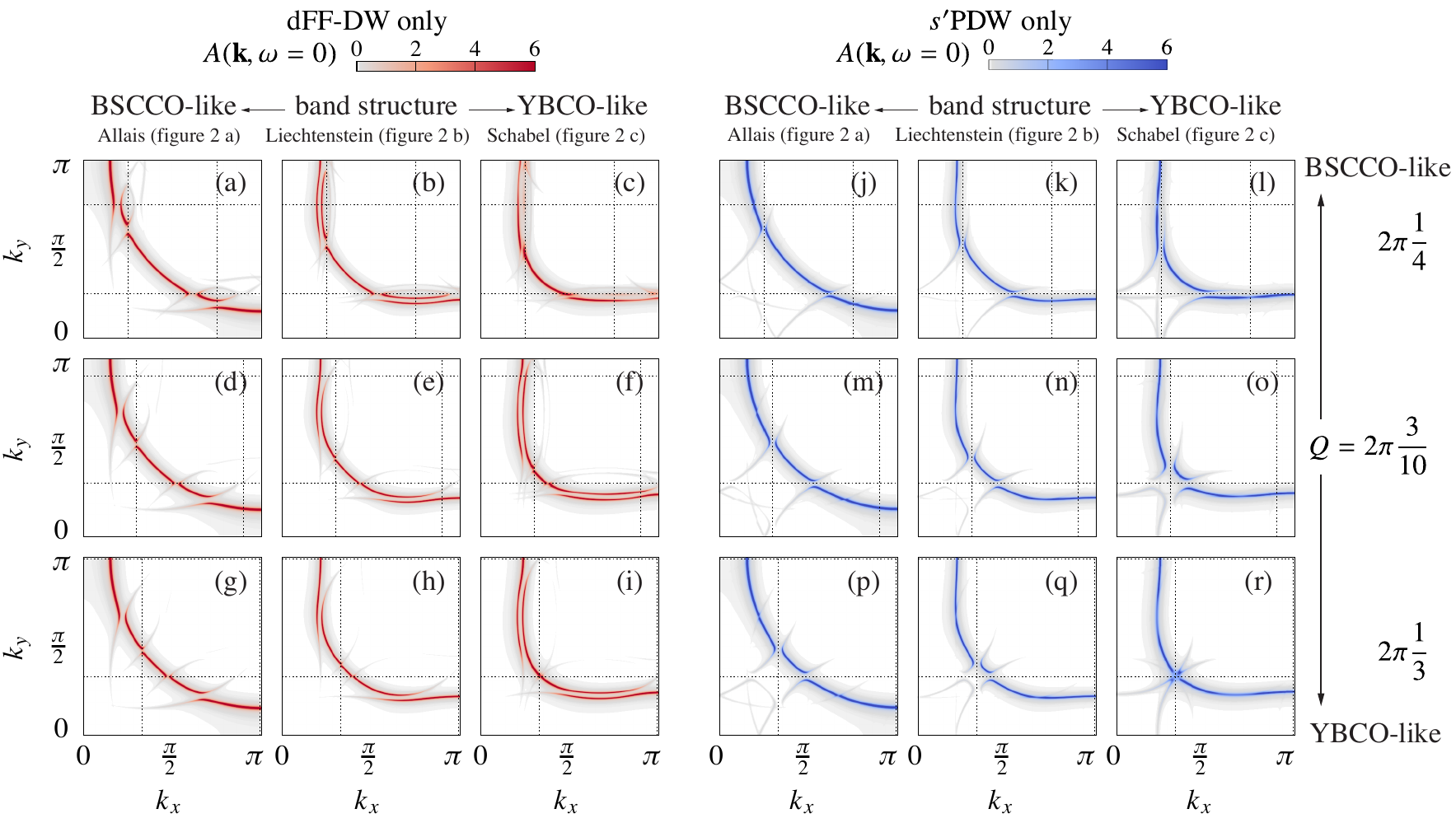} %[width=18cm]
\caption{
%Plots of the diagonal component of the momentum resolved spectral weight in the upper right quadrant of the original Brillouin zone. This illustrates the 
Great variability of the reconstructed Fermi-surfaces (FS): FS reconstructed by bidirectional bond-centered $d$-form factor density wave ($d$FF-DW with $t_Q=0.12t$) are illustrated in red (a)-(i), corresponding to red DOS of Fig.~\ref{figure_variability}, and FS reconstructed by $s'$-form factor pair density wave ($s'$PDW with $\Delta_Q=0.12t$) are in blue (j)-(r), corresponding to blue DOS of Fig.~\ref{figure_variability}. The dotted lines indicate the first order Bragg planes introduced by the DW periodicity. The reconstruction takes place on those Bragg planes, but also on higher-order Bragg planes (boundaries of the 2$^{\text{nd}}$, 3$^{\text{rd}}$ and successive reduced Brillouin zones), explaining why some gaps open away from dotted lines. Again, rows display FS for fixed wave vector and different band structures; columns display those for fixed band structures with different wave vectors. Lorentzian broadening is $\eta=0.02t$ with $t=250$ meV.}
\label{figure_mdc_variability}
\end{figure*}

\section{Results}\label{Sec:Results}

Fig.~\ref{figure_full_dos} shows typical densities of states (DOS) in the presence of a bidirectional $d$FF-DW alone---without the presence of $d$SC or $s'$PDW---for various magnitudes $t_Q$. The wave vectors shown are $Q=2\pi\tfrac{1}{3}$ and $Q=2\pi\tfrac{3}{10}$, with band structure from Liechtenstein (Fig.~\ref{band_structures}(b)). No pseudogap (PG) is obtained in the low energy DOS of such system. Instead, the actual gaps caused by the DWs appear high above and below the Fermi energy (around $\pm250$~meV). Although not shown on the figure, this result holds for the three band structures considered, the three wave vectors considered, and for all magnitudes of \mbox{$d$FF-DW}. Values higher than~$t_Q=0.28t$ display a broad structure at the Fermi energy. The finer details of this structure may be confused with a PG since two smaller peaks appear on each side of the Fermi energy. However, these two peaks move towards one another with increasing~$t_Q$ (see Fig.~\ref{figure_full_dos}, dotted lines), and this is inconsistent with the experimental fact that the size of the PG doesn't change when increasing the strength of $d$FF-DW (by decreasing temperature, for example).
Moreover, this structure appears only for $t_Q>0.28t$, whereas, according to experiments, charge density wave and superconductivity are expected to have comparable characteristic energies~\cite{chang_direct_2012} ($t_Q\approx\Delta_Q=0.12t=30$~meV in our model, yielding no PG in Fig.~\ref{figure_full_dos}).

The results of Fig.~\ref{figure_full_dos} are completely analogous to those for unidirectional dFF-DW (not shown). In the latter case, no gap is found near the Fermi level up to $t_Q=0.60t$. Above this value appears a peak structure that is similar but smaller than that of the bidirectional case. One notable exception, in the unidirectional case, is for the extreme value $t_Q=2.4t=600$~meV, in which case a 1200~meV wide partial gap caused by the DW reaches the Fermi level (not shown). Although such high values of $t_Q$ are unlikely for cuprates, we cover it in the discussion.

When pairing is added, the superconducting gap is very sensitive to both the wave vector $Q=2\pi\frac{q}{L}$ of the applied DWs, and the underlying band structure. Fig.~\ref{figure_variability} shows the low energy DOS for all 9 combinations of wave vector, $\frac{q}{L}=\frac{1}{4}$, $\frac{3}{10}$ and $\frac{1}{3}$, and band structure from Allais (Fig.~\ref{band_structures}(a)), Liechtenstein (Fig.~\ref{band_structures}(b)) and Schabel (Fig.~\ref{band_structures}~(c)) with mean fields of magnitudes $t_Q=0.12t$ (bidirectional $d$FF-DW), $\Delta_Q=0.12t$ (bidirectional $s'$PDW) and \mbox{$\Delta=0.12t$} ($d$SC), applied alone and combined (see key of Fig.~\ref{figure_variability}). The differences between each case demonstrate that band structure and DW's wave vector influence equally the deformation of the superconducting gap. The results are representative of those for other values of $\Delta$, $t_Q$ and $\Delta_Q$ ranging from $0.04t$ to $0.28t$ (not shown); values higher than $0.28$ were not investigated thoroughly. Results are perfectly analogous of those for unidirectional DWs (not shown).

A first observation for Fig.~\ref{figure_variability} is that, in the absence of $d$SC, again, no gap nor pseudogap appears near the Fermi energy for $d$FF-DW alone (red dotted lines). Similarily, $s'$PDW alone do not appreciably modify the low energy DOS of the normal state, and in that sense this is also a gapless state at the Fermi level, even if it is pairing. This occurs because of the finite pairing wave-vector. Hence, neither charge density wave, nor pair density waves, applied alone, for those wave vectors and band structures, resulted in a structure at the Fermi energy ($\omega=0$) that could be interpreted as a robust pseudogap. It is important to note that we did not adjust the wave vectors to perfect nesting since it does not seem to be the case experimentally~\cite{comin_charge_2014,mesaros_commensurate_2016}. An exception to the rule is when small gaps are apparent for $s'$PDW alone (Fig.~\ref{figure_variability}(b),~(f),~and~(i), dotted blue curves), which comes from scattering processes of higher order and correspond to a $\vec Q=0$ $s'$ homogeneous superconducting gap. 
Such a gap also appears when $d$FF-DW and $s'$PDW are applied together without $d$SC (Fig.~\ref{figure_variability}, dotted green curves). In that case, not only $s'$PDW scattering processes open a small gap, but also hybrid processes implying both $s'$PDW and $d$FF-DW. However, for these gaps to reach sizes comparable to the experimental PG, the mean fields $t_Q$ and $\Delta_Q$ must reach unlikely strengths, similar to those discussed above for the $d$FF-DW only case.

A second observation for Fig.~\ref{figure_variability} is that, although low-energy DOS curves in the absence of $d$SC are not much changed by DWs, the DOS in the presence of $d$SC are modified appreciably by the presence of $d$FF-DW and $s'$PDW. In some cases, the $d$SC gap exhibits major deformations even in the presence of weak density waves. This is true for $d$SC coexisting with $d$FF-DW (thin red curves),  for $d$SC coexisting with $s'$PDW (thicker blue curves), or for $d$SC coexisting with both (thickest green curves). Changes are diverse when compared to the bare $d$SC case (black curves), notable ones being : strongly reduced coherence peaks (Fig.~\ref{figure_variability}(b),~(c),~(f),~and~(i), thickest green curves) and displacement of coherence peaks, effectively changing the size of the gap (Fig.~\ref{figure_variability}(b),~(f),~and~(i), thin red curves). In other cases, the DWs have only a small effect (Fig.~\ref{figure_variability}(a),~(d),~(g),~and~(h) thin red and thickest green curves). This is highly reminiscent of the diversity of local conductance spectra observed in tunneling experiments~\cite{mcelroy_coincidence_2005, kohsaka_intrinsic_2007, machida_bipartite_2016} (especially in supplementary material of Ref.~\onlinecite{machida_bipartite_2016}). A very similar diversity of spectra is also obtained with unidirectional DWs, the main difference being that the various deformations and sub-gap structures are usually weaker. 

Our last and most important observation for Fig.~\ref{figure_variability} is that sub-gap structures, similar to those observed in the surrounding of vortex cores, appear mainly when $d$SC is accompanied by $s'$ pair density wave $s'$PDW (Fig.~\ref{figure_variability}(a),~(b),~(c),~(d),~(f),~and~(i), thick blue curves) without $d$FF-DW. When the three orders are present, sub-gap structures can also be obtained (Fig.~\ref{figure_variability}(c), thickest green curve) although they are more apparent with $s'$PDW stronger than $d$FF-DW (not shown). Additionally, when $d$FF-DW and $s'$PDW coexist without uniform $d$-wave superconductivity (Fig.~\ref{figure_variability}, dotted green curves), the small gap we obtain is reminiscent of what is measured directly inside the vortex core, where SC is mostly suppressed in the experiments~\cite{bruer_revisiting_2016,matsuba_anti-phase_2007, machida_bipartite_2016}.

To help understand the results for the density of states, Fig.~\ref{figure_mdc_variability} shows the Fermi surfaces (FS, $A(\vec k, \omega=0)$) for all nine combinations of wave vectors and band structures, for bidirectional $d$FF-DW with $t_Q=0.12t$ ((a)~to~(i), in red) and for bidirectional $s'$PDW with $\Delta_Q=0.12t$ ((j)~to~(r), in blue), without superconductivity.
As was seen in Fig.~\ref{figure_full_dos}, $d$FF-DW is gapless at the Fermi level, and therefore
no portion of the FS is completely gapped in Fig.~\ref{figure_mdc_variability}(a)~to~(i).
Surprisingly, this is also true for $s'$PDW: no portion of the FS is completely gapped in Fig.~\ref{figure_mdc_variability}(j)~to~(r), showing that $s'$PDW is also gapless at the Fermi level.
In all cases, the FS presents many hole and electron pockets, separated by the new Bragg planes imposed by the DWs. There is still spectral weight at most of the wave vectors of the original FS. 
For unidirectional DWs (not shown), the only difference is that only one direction of Bragg planes causes such pockets.
The shape and size of those pockets are extremely sensitive to the band structure and wave vector used.
Note that the FS shown are for DWs without $d$SC, and that the corresponding DOS (red and blue dotted curves of Fig.~\ref{figure_variability}) were left almost unchanged around the Fermi energy.

\section{Discussion}\label{Sec:Discussion}

We first discuss all of the preceding results. In a closing subsection, we briefly address the case where DWs appear on a pre-existing pseudogap.

\subsection{Density waves and sub-gap structures}

No gap appears in our low energy DOS without $d$SC because, in general, a DW alone does not open-up a gap at the Fermi surface. Contrary to the superconducting gap, a DW gap is not locked to the Fermi energy. In one dimension, for example, a DW of period $L$ simply splits the Brillouin zone in $L$, therefore separating the original band in $L$ new bands each containing $1/L$ of all the states. The $L-1$ gaps would thus appear at fixed fillings in the band, regardless of the energy. In two dimensions, for small DW magnitudes, the new bands do not have to split completely; they can overlap in $\omega$ space without touching in $\vec k$-space, and this yields a more complicated picture. Still, in the limit of very large DW magnitudes $t_Q$, the original band has to split completely into new bands containing a fixed fraction of the total number of states. This fraction is determined by the wave vectors of the DW at play. For example, the bidirectional $d$FF-DW of Fig.~\ref{figure_full_dos}(a), composed of two perpendicular density waves with $Q\equiv2\pi\tfrac{q}{L}=2\pi\tfrac{1}{3}$ should split the band in 9 (because the exact treatment requires a $L^2\times L^2$ matrix with $L=3$, providing 9 eigenvalues). Each of those bands occupies a $\vec k$-space volume equal to that of the reduced Brillouin zone, and therefore contains $1/9$ of the states. Consequently, full gaps should only appear at fillings that are multiples of $1/9$. The details of the splitting depend strongly on the symmetries of the system, on the original band structure, and on the DW's form factor. As one can see in Fig.~\ref{figure_full_dos}(a), for wave vector $Q=2\pi\tfrac{1}{3}$ the main gaps are exactly at fillings $1/3$ and $2/3$. The picture is similar for Fig.~\ref{figure_full_dos}(b), showing wave vector $Q=2\pi\tfrac{3}{10}$, which indicates that incommensurability plays a minor role. The hole doping used in this work  is $p=1/8=0.125$ relative to half-filling (1/2), and corresponds to an absolute filling of 7/16 and therefore, in our results, for wave vectors close to experimental ones, the Fermi level never lies in a DW gap, as seen in Fig.~\ref{figure_full_dos}.

This observation---the fact that a DW gap is constrained to appear at a wave-vector determined filling instead of a given energy---rules out most scenarios explaining the pseudogap with DW. The only density waves which would open a gap at the Fermi surface are those with wave vectors connecting flat segments of the Fermi surface, and they are inconsistent with combined photoemission, X-ray results~\cite{comin_charge_2014} and detailed analysis of scanning tunneling spectroscopy~\cite{mesaros_commensurate_2016}. Our conclusion therefore reinforces the idea that the PG is a different phenomenon. The present analysis, however, does not hold for more complex coupling in $\vec k$-space, for example Lee's Amperean pairing and so-called $2k_F$ scenarios~\cite{lee_amperean_2014, lee_cold-spots_2016, kloss_su2_2015, montiel_angle_2016}.

There is one additional notable exception to this rule: given sufficiently large $t_Q$ for a DW with wave-vectors connecting flat segments of the dispersion slightly above, or slightly below the Fermi level, a gap centered away from the Fermi surface could be large enough to deplete states at the Fermi level.
One example of such a gap is well known: it is the case of a cuprate-like dispersion gapped by a fairly large antiferromagnetic (AF) mean-field (shown in Fig.~\ref{figure_pg_scenario} and discussed in next section). At small hole doping, this system develops small electron pockets around $(\pm\pi,0)$ and $(0,\pm\pi)$, and small hole pockets around around $(\pm\frac{\pi}{2},\pm\frac{\pi}{2})$. Increasing the AF mean-field amplitude pushes the electron pockets away from the Fermi level, effectively gapping the antinodes. This effect can be considered as an antiferromagnetic pseudogap.
The only gap of this kind we found for $d$FF-DW was in the unidirectional case with an exaggerated $t_Q$ of $1.2t$. For smaller, reasonable values of $t_Q$, such a gap should be considered accidental; strictly speaking, the gap forms away from the Fermi surface (at an energy where the wave vector connects flat parts of the dispersion) and whether or not it reaches the Fermi level depends on the details of the underlying band structure and on the amplitude~$t_Q$.

Let's now turn to the densities of states in the presence of $d$SC.
A $d$-wave gap is anisotropic; the gapping energy changes with the direction of~$\vec k$. Thus, the DOS value at a given energy in the gap ---~i.e. the number of states at a given energy between the coherence peaks~--- is tied to the number of states contained along a specific direction in $\vec k$-space (the spectral weight). With a smooth FS, the spectral weight changes monotonically as a function of direction in $\vec k$-space, and therefore the gap is smooth as a function of energy: it is the well-known V-shaped gap of standard $d$-wave superconductors. On the other hand, for a $d$-wave gap built on a complex reconstructed FS (shown in Fig.~\ref{figure_mdc_variability}), the spectral weight is very irregular as a function of direction, and this is why we obtain an irregular gap as a function of energy.
In the end, only the anisotropy of the gap and the complexity of the bands onto which it opens are necessary to account for the complicated DOS we find. This last statement encompasses much beyond the systems explored here, as the next section suggests. 

In our computations, the final form of substructures are most sensitive to the factors determining the underlying complicated FS; here those were: (i) the nature of the reconstruction (whether it was $s'$PDW or $d$FF-DW), (ii) the wave vector of the DWs, and (iii) the initial band structure. Varying the value of the superconducting mean-field $\Delta$ did not have as much influence (not shown here).

Among all DOS deformations obtained, sub-gap structures like those of tunneling experiments agree better with our DOS reconstructed by PDW. This suggests that the PDW measured in BSCCO~\cite{hamidian_detection_2016} may be responsible for the substructures observed in cuprates. Moreover, as worked out by Agterberg and Garaud~\cite{agterberg_checkerboard_2015}, a vortex core can favor PDW through the suppression of $d$SC, so our result is consistent with the experimental fact that substructures are enhanced around vortex cores. Note, however, that the form of PDW used by Agterberg and Garaud's work is not the same as ours.

An alternative and intuitive explanation for the enhancement of PDW near vortices might be that the non-zero momentum $\vec Q$ carried by PDWs accommodates better the high currents expected near vortex cores. However, in our model, there is no vortex, and every $s'$PDW included are combinations of $+Q$ and $-Q$ causing no actual currents.

Let's now turn to the spectral functions displayed in Fig.~\ref{figure_mdc_variability}. Photoemission spectroscopy can access this information, in principle, but the rich structures obtained here was never reported in experiments, one possible explanation being that they are washed away by the short range nature of those order parameters \cite{lee_cold-spots_2016}. Nevertheless, Ref.~\onlinecite{allais_connecting_2014} showed that nodal hole pockets like the ones obtained here can account quantitatively for quantum oscillations measured in cuprates. Here, the great sensitivity of these pockets to the underlying band structure shows that such agreement necessitates very fine tuning. On one hand, if the wave vector, form factor and amplitude of the DW can be obtained from independent measurements, quantum oscillations will allow one to extract the original band structure with great accuracy. On the other hand, the sensitivity to band structure challenges the analysis of Ref.~\onlinecite{allais_connecting_2014}, since very different results would have been obtained had different band parameters been used.

The very careful observer might notice $x$-$y$ anisotropy in some of our $d$FF-DW Fermi-surfaces in Fig.~\ref{figure_mdc_variability} (of which the least subtle trace is found in the bottom left part of Fig.~\ref{figure_mdc_variability}(i)). This anisotropy is expected, as explained in the caption of Fig.~\ref{DWpatterns}, because of the form factor used here. This anisotropy might, moreover, cause even more complexity in the DOS, because the lost $C_4$ symmetry means that adjacent quadrants of $\vec k$-space could host reconstructed pockets that open and close at slightly different energies~$\omega$.

Finally, note that a perfect fit to experiments was not the aim here, neither for the DOS nor the spectral weight, since we focused on general conditions enabling sub-gap structures. As discussed, the results are extremely sensitive to the band structure used, and knowing that BSCCO and YBCO are in fact bi-layer compounds, and that modeling the bi-layer splitting is known to improve the fit to conductance spectra substantially~\cite{hoogenboom_modeling_2003}, we would suggest that bi-layer splitting be included to achieve better fits. Similarily, taking into account the spatial dependence of the Wannier functions seems to be very important for a quantitative fit~\cite{kreisel_interpretation_2015,choubey_incommensurate_2016}.

\subsection{When DWs occur over a pre-existing pseudogap}\label{Sec:WithPseudogap}

Since we showed that the PG is not caused by $d$FF-DW or $s'$PDW, a natural extension to our investigation is to consider what happens when DWs appear on top of an already pseudo-gapped dispersion rather than the non-interacting bands we used up to now. A particularly clear illustration of this idea is given in Ref.~\onlinecite{atkinson_charge_2015}, where an antiferromagnetic order is used to gap the antinode, leaving near the nodal position small hole pockets which nesting DWs wave vectors agree with experiments.

Most phenomenological theories of the pseudogap (PG) may be separated in two categories.

On the one hand, there are phenomenological PG inspired from the $d$SC gap; they open a particle-hole symmetric gap with a $d$-wave form factor at the Fermi level. An example of this is the pairing scattering model used to fit photoemission data in Refs.~\onlinecite{norman_phenomenology_1998} and~\onlinecite{kondo_point_2015}. For that kind of PG, the interplay with the FS reconstructed by DWs will be equivalent to that of the $d$SC cases presented in the two previous sections, and therefore, sub-gap structures similar to those discussed up to here can be expected.

On the other hand, there are phenomenological PGs having more in common with an antiferromagnetic (AF) gap; they split the dispersion in two, yielding a strongly particle-hole asymmetric gap. Examples of such theories include YRZ theory~\cite{yang_phenomenological_2006} and effective forms of fractionalized Fermi liquid (FL*)~theory~\cite{qi_effective_2010}. Ref.~\onlinecite{leblanc_signatures_2014} provides a good overview of the common characteristics of the gaps in these approximations, and Ref.~\onlinecite{borne_signature_2010} provides a detailed account of the exact behavior of the gap in YRZ theory.

\begin{figure}
\includegraphics{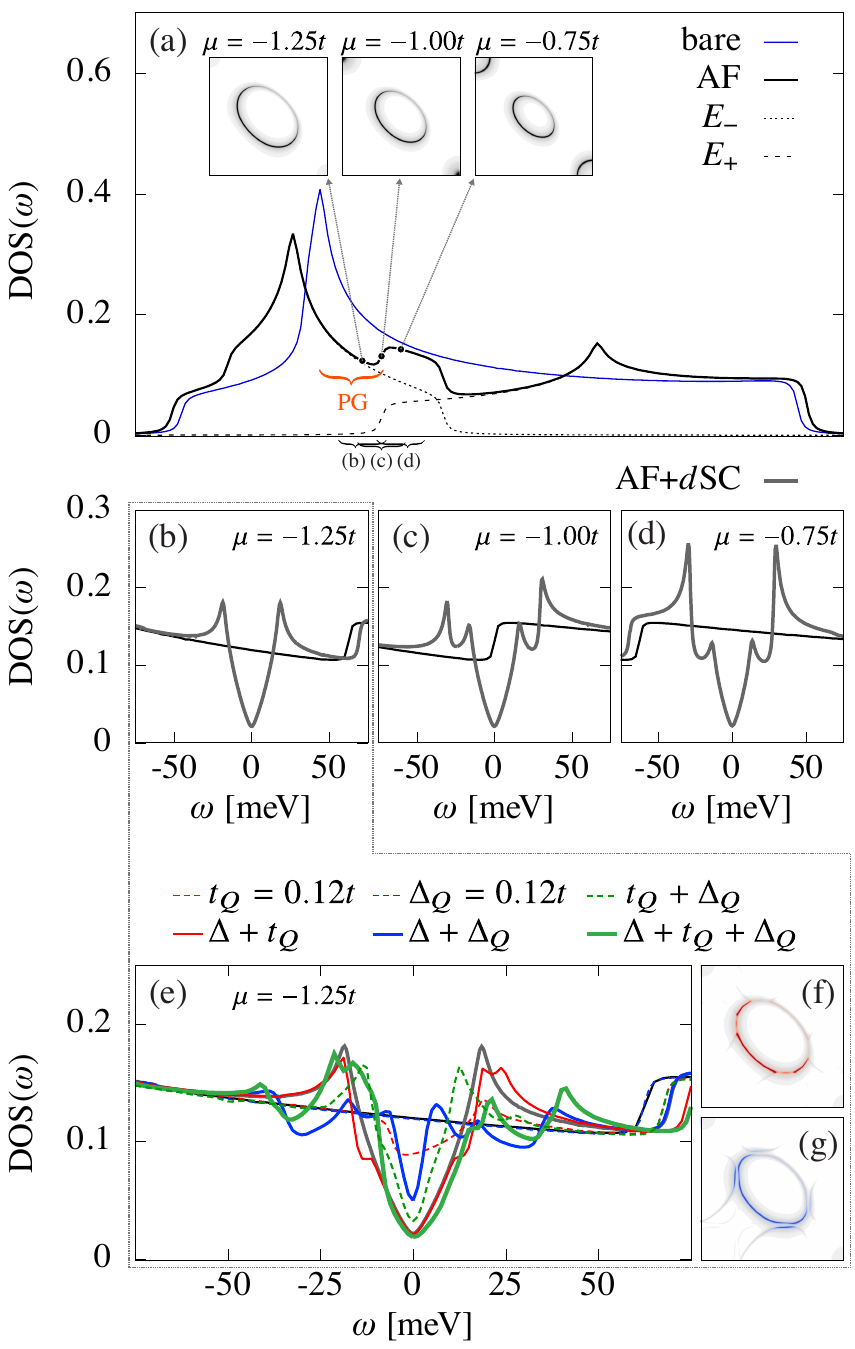}
\caption{Pseudogap simulated with antiferromagnetism (AF) and various possibilities of sub-gap structures caused by coexisting $d$SC, $d$FF-DW and $s'$PDW. (a) Density of states (DOS) for Liechtenstein band structure ($t'=-0.3$, $t''=0.2$) in thin blue, and with added AF mean-field of amplitude $0.5t$ in solid black. The three insets show the Fermi surface for chemical potential $\mu=-1.25t$, $-t$, and $-0.75t$, falling in the lower AF energy band ($E_-$, dotted), at the onset of the higher AF energy band ($E_+$, dashed). (b),(c),(d) For the same chemical potential values, enlargement around the Fermi level show the DOS for the AF mean-field alone (solid black) and coexisting with $d$SC (thicker gray). (e) For $\mu=-1.25t$ only, DOS near the Fermi level for the AF mean-field (solid black), and coexisting with the following: $d$SC (thicker gray line), $d$FF-DW (dotted red), $s'$PDW (dotted blue), $d$FF-DW and $s'$PDW (dotted green), $d$SC and $d$FF-DW (solid red), $d$SC and $s'$PDW (solid blue), $d$SC and $d$FF-DW and $s'$PDW (solid green). Spectral weight at the Fermi surface (f) for dFF-DW on an AF pseudogap and (g) for s’PDW on an AF pseudogap. Color scales for (f) and (g) are the same as in Fig.~\ref{figure_mdc_variability}}
\label{figure_pg_scenario}
\end{figure}

To understand how DWs may lead to sub-gap structures in those AF-like theories, we consider Fig.~\ref{figure_pg_scenario}, where a simple AF mean-field was added to the Hamiltonian in order to simulate a pseudogap, using band structure from Liechtenstein ($t'=-0.3$, $t''=0.2$).
The AF mean-field follows equation~\eqref{eqdffdw}, with \mbox{$t_{(\pi,\pi),\vec 0}=0.5t$} for spin up and \mbox{$t_{(\pi,\pi),\vec 0}=-0.5t$} for spin down. The resulting AF density of state (DOS) is compared with the bare DOS in Fig.~\ref{figure_pg_scenario}(a). Separate lines for the DOS of the lower AF band ($E_-$) and the higher AF band ($E_+$) illustrate how they overlap to yield a particle-hole asymmetric partial gap, identified here as the PG.
As shown by the Fermi surfaces in the insets, the higher energy edge of the AF pseudogap corresponds to the bottom of the higher AF band (consisting of electron pockets). Ref.~\onlinecite{leblanc_signatures_2014} and~\onlinecite{borne_signature_2010} show similar behavior for YRZ theory.

Strictly speaking, the PG identified in Fig.~\ref{figure_pg_scenario}(a) would become a full DW gap at half-filling if the mean-field was stronger (because AF is a DW). In Fig.~\ref{figure_pg_scenario}(a), however, the mean-field is not strong enough to split the band completely, but, as explained in the previous section, it is strong enough to partially gap the Femi-level given the right chemical potential.
%We could vary the mean-field with doping so that the gap appears at the Fermi level for all dopings, as done in YRZ theory for $\Delta_{PG}(p)$, but here we fix its value at $0.5t$ for simplicity.

\begin{figure}
\includegraphics{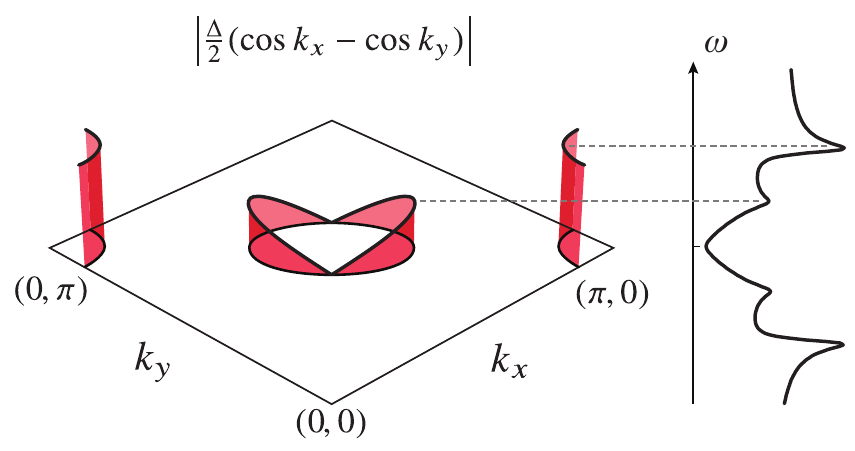}
\caption{Illustration of how multiple pockets on the Fermi surface cause additional coherence peaks in the $d$-wave gap. On the left, the amplitude of the $d$-wave gap $\vert\tfrac{\Delta}{2}(\cos k_x -\cos k_y)\vert$ is shown for all $\vec k_F$ vectors on the Fermi surface. On the right, the density of states (DOS) is shown at the corresponding energy. This picture is for the two-band antiferromagnetic system of Fig.~\ref{figure_pg_scenario}(d), for $\mu=-0.75t$, but the explanation holds for other reconstructed band structures.}
\label{explanation}
\end{figure}

Let us first consider how the $d$SC gap behaves when added to this system. In Fig.~\ref{figure_pg_scenario}(b), at chemical potential $\mu=-1.25t$, the Fermi level is below the bottom of the higher AF band, and the $d$SC gap appears inside the AF pseudogap. This $d$SC gap does not display sub-structures and, instead, could be interpreted as a sub-structure of the pseudogap, as in Ref.~\onlinecite{pushp_extending_2009} and \onlinecite{he_fermi_2014}.
In Fig.~\ref{figure_pg_scenario}(c) and (d), at chemical potentials $\mu=-t$ and $-0.75t$, the electron pockets of the higher AF band are brought to the Fermi level. The $d$SC gap then appears on the edge (Fig.~\ref{figure_pg_scenario}(c)) or outside (Fig.~\ref{figure_pg_scenario}(d)) of the PG. In these cases, the $d$SC gap displays new sub-structures, similar to those reported in Ref.~\onlinecite{andersen_extinction_2009}.
The additional coherence peaks appear at higher energy than those of Fig.~\ref{figure_pg_scenario}(b). Indeed, the new electron pockets are centered at $(\pi,0)$ and $(0,\pi)$ where the $d$-wave gap energy $\tfrac{\Delta}{2}(\cos k_x-\cos k_y)$ is at its maximum. Hence, the spectral weight coming from these new pockets is gapped to a higher energy coherence peak than for the hole pocket, as schematized in Fig.~\ref{explanation}. 
The new peaks are therefore a direct consequence of the electron pockets getting in range of the $d$SC gap. This simple AF example provides a clear illustration of what we explained in the previous section: multiple DWs pockets on the Fermi surface lead to sub-structures in the $d$SC gap.

Sub-structures perfectly analogous to those of Fig.~\ref{figure_pg_scenario}(c) and~(d) have been obtained previously in YRZ theory, when $d$SC is present (see Fig.~6(d) of Ref.~\onlinecite{borne_signature_2010}). In YRZ theory, the gapping parameter $\Delta_{PG}(p)$ decreases with doping, and the anti-nodal electron pockets causing these sub-structures only appear at very high doping, for small $\Delta_{PG}(p)$. In experiments, sub-gap structures are rather found in underdoped samples, when the PG is strong. Hence, the sub-gap structures seen in YRZ theory and the AF model of Fig.~\ref{figure_pg_scenario}(c) and (d) are not likely to be those seen in experiments.

We then turn to the case of interest, illustrated in Fig.~\ref{figure_pg_scenario}(e), (f) and (g), when the DWs of previous sections nest on the AF hole pocket. We have chosen the AF mean-field value ($0.5t$) and doping ($\mu=-1.25t$, as in Fig.~\ref{figure_pg_scenario}(b)) so that the nesting condition is fulfilled when the AF electron pockets are absent\footnote{The corresponding doping (approximately $p=0.2$) is not really relevant; we could tune the band structure, the AF mean-field and/or the DWs wave vector (here $\frac{q}{L}=\frac{1}{3}$), to fit the experiments, as in Ref.~\onlinecite{atkinson_charge_2015}.}, and thus without preexisting sub-structures in the $d$SC~gap.

Fig.~\ref{figure_pg_scenario}(e) demonstrates that nesting DWs on the hole pocket alone cause another set of deformations in the $d$SC gap, distinct from those just considered in Fig.~\ref{figure_pg_scenario}(c) and~(d). These new sub-structures are analogous to those discussed in the previous sections. Indeed, from the standpoint of $d$SC, only low-energy states are affected, and therefore the AF hole pocket is equivalent to another single band system (electrons pockets are far from the Fermi level here, since $\mu=-1.25t$). DWs split the AF hole pocket in more pockets, and thus the $d$SC gap is deformed, as in the previous section. However, sub-structures now appear inside a $d$SC gap that is already inside a PG, so this yield a three-gap picture more complicated than the two gaps usually reported in experiments.

That DWs are \emph{nesting} in the hole pocket also introduces some new features. First, $d$FF-DW has a gap at the Fermi level even without $d$SC. This is seen on the red dotted DOS of Fig.~\ref{figure_pg_scenario}(e) and the FS of Fig.~\ref{figure_pg_scenario}(f). This DW gap inside the AF gap could also be interpreted as a two-gap structure in its own right. Second, contrary to $d$FF-DW, in the absence of dSC a nesting $s'$PDW does not open a gap at the Fermi level. Instead, it simply splits the edge of the hole pocket, as seen on Fig.~\ref{figure_pg_scenario}(g). 
\footnote{This is due to the nature of the $s'$PDW, which couples the band energy $\xi(\vec k)$ with $-\xi(\vec k+\vec Q)$.}
As a consequence of this splitting, the deformations of the $d$SC gap caused by $s'$PDW are a lot stronger than for non-nesting $s'$PDW (seen in Fig.~\ref{figure_variability}). Nevertheless, they still compare favorably to experimental spectra seen very close to vortex cores~\cite{bruer_revisiting_2016}.

To sum up, the two kinds of $T=0$ phenomenological PG considered here reflect well the dichotomy between ``preformed pairs'' and ``distinct phase'' scenarios~\cite{kordyuk_pseudogap_2015}.
In the first kind, the PG resembles a $d$SC gap, and sub-structures similar to those seen in experiments would be the result of coexisting $s'$PDW. In this case the comparison with experiments is striking (especially with the supplementary material of Ref.~\onlinecite{machida_bipartite_2016}).
In the second kind, where DWs nest on an already formed AF-like hole pocket, sub-gap structures could be obtained with (i) a simple $d$SC gap inside the already formed PG, (ii) a nesting $d$FF-DW gap inside an already formed PG or (iii) $d$SC coexisting with DWs inside the already formed PG. But in all cases where the AF pseudogap is present,  the comparison with experiments is less convincing. 
In particular, coexisting $d$SC and DWs inside an already formed pseudogap yielded more complicated gap structure than what is usually seen in experiments. 
%Yet, the correlation in position of DWs with experimental sub-gap structures makes the link between the two hard to dismiss.
Therefore, within mean-field type theories, the pseudogap as a distinct phenomenon (here AF order), seems superfluous, and even detrimental, to the agreement with experiment.

\section{Conclusion}

We found that sub-gap structures like those seen near vortex cores, and sometimes reported at zero-field, appear as deformations of the $d$SC gap when it coexists with DWs, especially pair density waves.
Arguably, the pair density wave is enhanced and sub-gap structures appear more clearly in the vicinity of a vortex core (where checkerboard DWs were first discovered), in high field, near impurities, or simply in extreme cases of inhomogeneity. Our results thus suggest that the PDW measured in BSCCO~\cite{hamidian_detection_2016} is responsible for the substructures observed in STM.

We also demonstrated, in the mean-field approximation and for reasonable one-band models, that neither $d$FF-DW nor $s'$PDW alone could be responsible for the opening of the pseudogap in the DOS in the absence of nesting. Therefore, this work reinforces the idea that the pseudogap is a distinct phenomenon.

Instead of causing a pseudogap, DWs cause various kinds of deformations of the $d$SC gap, comparable to the variability of local tunneling spectra. 
Those deformations of the DOS depend as strongly on the band-structures involved than they do on the wave vector of the~DW.

For a simple PG model where DWs nest the nodal hole pockets of an antiferromagnet acting as proxy for the pseudogap, we demonstrated similar deformations for the $d$SC gap.
However, a two-band model like the antiferromagnet seemed detrimental to the agreement with experiments.
%, underlining the pertinence of more subtle theories of the PG\cite{hayward_angular_2014, lee_amperean_2014, kloss_su2_2015, lee_cold-spots_2016, fratino_organizing_2016}.

To conclude, let us mention how further practical constraints on theories emerge from the idea that sub-gap structures are the consequence of DWs. The behavior of experimental sub-gap structures with position is very specific~\cite{matsuba_anti-phase_2007, machida_bipartite_2016}, for example their characteristic energy does not change much when approaching a vortex core. On the other hand, our calculated sub-gap structures depend strongly on the parameters at play, namely the DW wave vector, the magnitudes $\Delta$, $t_Q$ and $\Delta_Q$, and band-structure. Hence, in order to achieve quantitative agreement with experiments, a theory of cuprates would have to predict a delicate balance between those parameters, or their equivalent. Such fine tuning would be very surprising coming from a simple mix of competing orders. This work therefore provides a good justification, beyond simple taste for unified theories, to search for the common origin of intertwined orders in cuprates~\cite{atkinson_charge_2015, fradkin_colloquium_2015, kloss_su2_2015, hayward_angular_2014, fratino_organizing_2016, tu_genesis_2016,choubey_incommensurate_2016}.

\section{Acknowledgments}
We acknowledge W.A.~Atkinson, P.~Choubey, S.~Edkins, J.C.S.~Davis, A.~Foley, R.~Frésard, P.~Hirschfeld, J.~Hoffman, R. Nourafkan, C.~P\'epin, S.~Sachdev, and L.~Taillefer for discussions. This work was partially supported by the Natural Sciences and Engineering Research Council (Canada) under grant RGPIN-2014-04584, the Fonds Nature et Technologie (Qu\'ebec) and the Research Chair on the Theory of Quantum Materials (A.-M.S.T.). Simulations were performed on computers provided by the Canadian Foundation for Innovation, the Minist\`ere de l’\'Education des Loisirs et du Sport (Qu\'ebec), Calcul Qu\'ebec, and Compute Canada.

%merlin.mbs apsrev4-1.bst 2010-07-25 4.21a (PWD, AO, DPC) hacked
%Control: key (0)
%Control: author (8) initials jnrlst
%Control: editor formatted (1) identically to author
%Control: production of article title (-1) disabled
%Control: page (0) single
%Control: year (1) truncated
%Control: production of eprint (0) enabled
%

%\bibliography{main}

\begin{thebibliography}{76}%
\makeatletter
\providecommand \@ifxundefined [1]{%
 \@ifx{#1\undefined}
}%
\providecommand \@ifnum [1]{%
 \ifnum #1\expandafter \@firstoftwo
 \else \expandafter \@secondoftwo
 \fi
}%
\providecommand \@ifx [1]{%
 \ifx #1\expandafter \@firstoftwo
 \else \expandafter \@secondoftwo
 \fi
}%
\providecommand \natexlab [1]{#1}%
\providecommand \enquote  [1]{``#1''}%
\providecommand \bibnamefont  [1]{#1}%
\providecommand \bibfnamefont [1]{#1}%
\providecommand \citenamefont [1]{#1}%
\providecommand \href@noop [0]{\@secondoftwo}%
\providecommand \href [0]{\begingroup \@sanitize@url \@href}%
\providecommand \@href[1]{\@@startlink{#1}\@@href}%
\providecommand \@@href[1]{\endgroup#1\@@endlink}%
\providecommand \@sanitize@url [0]{\catcode `\\12\catcode `\$12\catcode
  `\&12\catcode `\#12\catcode `\^12\catcode `\_12\catcode `\%12\relax}%
\providecommand \@@startlink[1]{}%
\providecommand \@@endlink[0]{}%
\providecommand \url  [0]{\begingroup\@sanitize@url \@url }%
\providecommand \@url [1]{\endgroup\@href {#1}{\urlprefix }}%
\providecommand \urlprefix  [0]{URL }%
\providecommand \Eprint [0]{\href }%
\providecommand \doibase [0]{http://dx.doi.org/}%
\providecommand \selectlanguage [0]{\@gobble}%
\providecommand \bibinfo  [0]{\@secondoftwo}%
\providecommand \bibfield  [0]{\@secondoftwo}%
\providecommand \translation [1]{[#1]}%
\providecommand \BibitemOpen [0]{}%
\providecommand \bibitemStop [0]{}%
\providecommand \bibitemNoStop [0]{.\EOS\space}%
\providecommand \EOS [0]{\spacefactor3000\relax}%
\providecommand \BibitemShut  [1]{\csname bibitem#1\endcsname}%
\let\auto@bib@innerbib\@empty
%</preamble>
\bibitem [{\citenamefont {Ghiringhelli}\ \emph {et~al.}(2012)\citenamefont
  {Ghiringhelli}, \citenamefont {Tacon}, \citenamefont {Minola}, \citenamefont
  {Blanco-Canosa}, \citenamefont {Mazzoli}, \citenamefont {Brookes},
  \citenamefont {Luca}, \citenamefont {Frano}, \citenamefont {Hawthorn},
  \citenamefont {He}, \citenamefont {Loew}, \citenamefont {Sala}, \citenamefont
  {Peets}, \citenamefont {Salluzzo}, \citenamefont {Schierle}, \citenamefont
  {Sutarto}, \citenamefont {Sawatzky}, \citenamefont {Weschke}, \citenamefont
  {Keimer},\ and\ \citenamefont {Braicovich}}]{ghiringhelli_long-range_2012}%
  \BibitemOpen
  \bibfield  {author} {\bibinfo {author} {\bibfnamefont {G.}~\bibnamefont
  {Ghiringhelli}}, \bibinfo {author} {\bibfnamefont {M.~L.}\ \bibnamefont
  {Tacon}}, \bibinfo {author} {\bibfnamefont {M.}~\bibnamefont {Minola}},
  \bibinfo {author} {\bibfnamefont {S.}~\bibnamefont {Blanco-Canosa}}, \bibinfo
  {author} {\bibfnamefont {C.}~\bibnamefont {Mazzoli}}, \bibinfo {author}
  {\bibfnamefont {N.~B.}\ \bibnamefont {Brookes}}, \bibinfo {author}
  {\bibfnamefont {G.~M.~D.}\ \bibnamefont {Luca}}, \bibinfo {author}
  {\bibfnamefont {A.}~\bibnamefont {Frano}}, \bibinfo {author} {\bibfnamefont
  {D.~G.}\ \bibnamefont {Hawthorn}}, \bibinfo {author} {\bibfnamefont
  {F.}~\bibnamefont {He}}, \bibinfo {author} {\bibfnamefont {T.}~\bibnamefont
  {Loew}}, \bibinfo {author} {\bibfnamefont {M.~M.}\ \bibnamefont {Sala}},
  \bibinfo {author} {\bibfnamefont {D.~C.}\ \bibnamefont {Peets}}, \bibinfo
  {author} {\bibfnamefont {M.}~\bibnamefont {Salluzzo}}, \bibinfo {author}
  {\bibfnamefont {E.}~\bibnamefont {Schierle}}, \bibinfo {author}
  {\bibfnamefont {R.}~\bibnamefont {Sutarto}}, \bibinfo {author} {\bibfnamefont
  {G.~A.}\ \bibnamefont {Sawatzky}}, \bibinfo {author} {\bibfnamefont
  {E.}~\bibnamefont {Weschke}}, \bibinfo {author} {\bibfnamefont
  {B.}~\bibnamefont {Keimer}}, \ and\ \bibinfo {author} {\bibfnamefont
  {L.}~\bibnamefont {Braicovich}},\ }\href {\doibase 10.1126/science.1223532}
  {\bibfield  {journal} {\bibinfo  {journal} {Science}\ }\textbf {\bibinfo
  {volume} {337}},\ \bibinfo {pages} {821} (\bibinfo {year}
  {2012})}\BibitemShut {NoStop}%
\bibitem [{\citenamefont {Comin}\ and\ \citenamefont
  {Damascelli}(2016)}]{comin_resonant_2016}%
  \BibitemOpen
  \bibfield  {author} {\bibinfo {author} {\bibfnamefont {R.}~\bibnamefont
  {Comin}}\ and\ \bibinfo {author} {\bibfnamefont {A.}~\bibnamefont
  {Damascelli}},\ }\href {\doibase 10.1146/annurev-conmatphys-031115-011401}
  {\bibfield  {journal} {\bibinfo  {journal} {Annual Review of Condensed Matter
  Physics}\ }\textbf {\bibinfo {volume} {7}},\ \bibinfo {pages} {369} (\bibinfo
  {year} {2016})}\BibitemShut {NoStop}%
\bibitem [{\citenamefont {Comin}\ \emph {et~al.}(2014)\citenamefont {Comin},
  \citenamefont {Frano}, \citenamefont {Yee}, \citenamefont {Yoshida},
  \citenamefont {Eisaki}, \citenamefont {Schierle}, \citenamefont {Weschke},
  \citenamefont {Sutarto}, \citenamefont {He}, \citenamefont {Soumyanarayanan},
  \citenamefont {He}, \citenamefont {Tacon}, \citenamefont {Elfimov},
  \citenamefont {Hoffman}, \citenamefont {Sawatzky}, \citenamefont {Keimer},\
  and\ \citenamefont {Damascelli}}]{comin_charge_2014}%
  \BibitemOpen
  \bibfield  {author} {\bibinfo {author} {\bibfnamefont {R.}~\bibnamefont
  {Comin}}, \bibinfo {author} {\bibfnamefont {A.}~\bibnamefont {Frano}},
  \bibinfo {author} {\bibfnamefont {M.~M.}\ \bibnamefont {Yee}}, \bibinfo
  {author} {\bibfnamefont {Y.}~\bibnamefont {Yoshida}}, \bibinfo {author}
  {\bibfnamefont {H.}~\bibnamefont {Eisaki}}, \bibinfo {author} {\bibfnamefont
  {E.}~\bibnamefont {Schierle}}, \bibinfo {author} {\bibfnamefont
  {E.}~\bibnamefont {Weschke}}, \bibinfo {author} {\bibfnamefont
  {R.}~\bibnamefont {Sutarto}}, \bibinfo {author} {\bibfnamefont
  {F.}~\bibnamefont {He}}, \bibinfo {author} {\bibfnamefont {A.}~\bibnamefont
  {Soumyanarayanan}}, \bibinfo {author} {\bibfnamefont {Y.}~\bibnamefont {He}},
  \bibinfo {author} {\bibfnamefont {M.~L.}\ \bibnamefont {Tacon}}, \bibinfo
  {author} {\bibfnamefont {I.~S.}\ \bibnamefont {Elfimov}}, \bibinfo {author}
  {\bibfnamefont {J.~E.}\ \bibnamefont {Hoffman}}, \bibinfo {author}
  {\bibfnamefont {G.~A.}\ \bibnamefont {Sawatzky}}, \bibinfo {author}
  {\bibfnamefont {B.}~\bibnamefont {Keimer}}, \ and\ \bibinfo {author}
  {\bibfnamefont {A.}~\bibnamefont {Damascelli}},\ }\href {\doibase
  10.1126/science.1242996} {\bibfield  {journal} {\bibinfo  {journal}
  {Science}\ }\textbf {\bibinfo {volume} {343}},\ \bibinfo {pages} {390}
  (\bibinfo {year} {2014})}\BibitemShut {NoStop}%
\bibitem [{\citenamefont {Alloul}(2014)}]{alloul_what_2014}%
  \BibitemOpen
  \bibfield  {author} {\bibinfo {author} {\bibfnamefont {H.}~\bibnamefont
  {Alloul}},\ }\href {\doibase 10.1016/j.crhy.2014.02.007} {\bibfield
  {journal} {\bibinfo  {journal} {Comptes Rendus Physique}\ }\textbf {\bibinfo
  {volume} {15}},\ \bibinfo {pages} {519} (\bibinfo {year} {2014})}\BibitemShut
  {NoStop}%
\bibitem [{\citenamefont {Atkinson}\ \emph {et~al.}(2015)\citenamefont
  {Atkinson}, \citenamefont {Kampf},\ and\ \citenamefont
  {Bulut}}]{atkinson_charge_2015}%
  \BibitemOpen
  \bibfield  {author} {\bibinfo {author} {\bibfnamefont {W.~A.}\ \bibnamefont
  {Atkinson}}, \bibinfo {author} {\bibfnamefont {A.~P.}\ \bibnamefont {Kampf}},
  \ and\ \bibinfo {author} {\bibfnamefont {S.}~\bibnamefont {Bulut}},\ }\href
  {\doibase 10.1088/1367-2630/17/1/013025} {\bibfield  {journal} {\bibinfo
  {journal} {New Journal of Physics}\ }\textbf {\bibinfo {volume} {17}},\
  \bibinfo {pages} {013025} (\bibinfo {year} {2015})}\BibitemShut {NoStop}%
\bibitem [{\citenamefont {Badoux}\ \emph {et~al.}(2016)\citenamefont {Badoux},
  \citenamefont {Tabis}, \citenamefont {Laliberté}, \citenamefont
  {Grissonnanche}, \citenamefont {Vignolle}, \citenamefont {Vignolles},
  \citenamefont {Béard}, \citenamefont {Bonn}, \citenamefont {Hardy},
  \citenamefont {Liang}, \citenamefont {Doiron-Leyraud}, \citenamefont
  {Taillefer},\ and\ \citenamefont {Proust}}]{badoux_change_2016}%
  \BibitemOpen
  \bibfield  {author} {\bibinfo {author} {\bibfnamefont {S.}~\bibnamefont
  {Badoux}}, \bibinfo {author} {\bibfnamefont {W.}~\bibnamefont {Tabis}},
  \bibinfo {author} {\bibfnamefont {F.}~\bibnamefont {Laliberté}}, \bibinfo
  {author} {\bibfnamefont {G.}~\bibnamefont {Grissonnanche}}, \bibinfo {author}
  {\bibfnamefont {B.}~\bibnamefont {Vignolle}}, \bibinfo {author}
  {\bibfnamefont {D.}~\bibnamefont {Vignolles}}, \bibinfo {author}
  {\bibfnamefont {J.}~\bibnamefont {Béard}}, \bibinfo {author} {\bibfnamefont
  {D.~A.}\ \bibnamefont {Bonn}}, \bibinfo {author} {\bibfnamefont {W.~N.}\
  \bibnamefont {Hardy}}, \bibinfo {author} {\bibfnamefont {R.}~\bibnamefont
  {Liang}}, \bibinfo {author} {\bibfnamefont {N.}~\bibnamefont
  {Doiron-Leyraud}}, \bibinfo {author} {\bibfnamefont {L.}~\bibnamefont
  {Taillefer}}, \ and\ \bibinfo {author} {\bibfnamefont {C.}~\bibnamefont
  {Proust}},\ }\href {\doibase 10.1038/nature16983} {\bibfield  {journal}
  {\bibinfo  {journal} {Nature}\ }\textbf {\bibinfo {volume} {531}},\ \bibinfo
  {pages} {210} (\bibinfo {year} {2016})}\BibitemShut {NoStop}%
\bibitem [{\citenamefont {He}\ \emph {et~al.}(2014)\citenamefont {He},
  \citenamefont {Yin}, \citenamefont {Zech}, \citenamefont {Soumyanarayanan},
  \citenamefont {Yee}, \citenamefont {Williams}, \citenamefont {Boyer},
  \citenamefont {Chatterjee}, \citenamefont {Wise}, \citenamefont {Zeljkovic},
  \citenamefont {Kondo}, \citenamefont {Takeuchi}, \citenamefont {Ikuta},
  \citenamefont {Mistark}, \citenamefont {Markiewicz}, \citenamefont {Bansil},
  \citenamefont {Sachdev}, \citenamefont {Hudson},\ and\ \citenamefont
  {Hoffman}}]{he_fermi_2014}%
  \BibitemOpen
  \bibfield  {author} {\bibinfo {author} {\bibfnamefont {Y.}~\bibnamefont
  {He}}, \bibinfo {author} {\bibfnamefont {Y.}~\bibnamefont {Yin}}, \bibinfo
  {author} {\bibfnamefont {M.}~\bibnamefont {Zech}}, \bibinfo {author}
  {\bibfnamefont {A.}~\bibnamefont {Soumyanarayanan}}, \bibinfo {author}
  {\bibfnamefont {M.~M.}\ \bibnamefont {Yee}}, \bibinfo {author} {\bibfnamefont
  {T.}~\bibnamefont {Williams}}, \bibinfo {author} {\bibfnamefont {M.~C.}\
  \bibnamefont {Boyer}}, \bibinfo {author} {\bibfnamefont {K.}~\bibnamefont
  {Chatterjee}}, \bibinfo {author} {\bibfnamefont {W.~D.}\ \bibnamefont
  {Wise}}, \bibinfo {author} {\bibfnamefont {I.}~\bibnamefont {Zeljkovic}},
  \bibinfo {author} {\bibfnamefont {T.}~\bibnamefont {Kondo}}, \bibinfo
  {author} {\bibfnamefont {T.}~\bibnamefont {Takeuchi}}, \bibinfo {author}
  {\bibfnamefont {H.}~\bibnamefont {Ikuta}}, \bibinfo {author} {\bibfnamefont
  {P.}~\bibnamefont {Mistark}}, \bibinfo {author} {\bibfnamefont {R.~S.}\
  \bibnamefont {Markiewicz}}, \bibinfo {author} {\bibfnamefont
  {A.}~\bibnamefont {Bansil}}, \bibinfo {author} {\bibfnamefont
  {S.}~\bibnamefont {Sachdev}}, \bibinfo {author} {\bibfnamefont {E.~W.}\
  \bibnamefont {Hudson}}, \ and\ \bibinfo {author} {\bibfnamefont {J.~E.}\
  \bibnamefont {Hoffman}},\ }\href {\doibase 10.1126/science.1248221}
  {\bibfield  {journal} {\bibinfo  {journal} {Science}\ }\textbf {\bibinfo
  {volume} {344}},\ \bibinfo {pages} {608} (\bibinfo {year}
  {2014})}\BibitemShut {NoStop}%
\bibitem [{\citenamefont {Hamidian}\ \emph
  {et~al.}(2016{\natexlab{a}})\citenamefont {Hamidian}, \citenamefont {Edkins},
  \citenamefont {Kim}, \citenamefont {Davis}, \citenamefont {Mackenzie},
  \citenamefont {Eisaki}, \citenamefont {Uchida}, \citenamefont {Lawler},
  \citenamefont {Kim}, \citenamefont {Sachdev},\ and\ \citenamefont
  {Fujita}}]{hamidian_atomic-scale_2016}%
  \BibitemOpen
  \bibfield  {author} {\bibinfo {author} {\bibfnamefont {M.~H.}\ \bibnamefont
  {Hamidian}}, \bibinfo {author} {\bibfnamefont {S.~D.}\ \bibnamefont
  {Edkins}}, \bibinfo {author} {\bibfnamefont {C.~K.}\ \bibnamefont {Kim}},
  \bibinfo {author} {\bibfnamefont {J.~C.}\ \bibnamefont {Davis}}, \bibinfo
  {author} {\bibfnamefont {A.~P.}\ \bibnamefont {Mackenzie}}, \bibinfo {author}
  {\bibfnamefont {H.}~\bibnamefont {Eisaki}}, \bibinfo {author} {\bibfnamefont
  {S.}~\bibnamefont {Uchida}}, \bibinfo {author} {\bibfnamefont {M.~J.}\
  \bibnamefont {Lawler}}, \bibinfo {author} {\bibfnamefont {E.-A.}\
  \bibnamefont {Kim}}, \bibinfo {author} {\bibfnamefont {S.}~\bibnamefont
  {Sachdev}}, \ and\ \bibinfo {author} {\bibfnamefont {K.}~\bibnamefont
  {Fujita}},\ }\href {\doibase 10.1038/nphys3519} {\bibfield  {journal}
  {\bibinfo  {journal} {Nature Physics}\ }\textbf {\bibinfo {volume} {12}},\
  \bibinfo {pages} {150} (\bibinfo {year} {2016}{\natexlab{a}})}\BibitemShut
  {NoStop}%
\bibitem [{\citenamefont {Hoffman}\ \emph {et~al.}(2002)\citenamefont
  {Hoffman}, \citenamefont {Hudson}, \citenamefont {Lang}, \citenamefont
  {Madhavan}, \citenamefont {Eisaki}, \citenamefont {Uchida},\ and\
  \citenamefont {Davis}}]{hoffman_four_2002}%
  \BibitemOpen
  \bibfield  {author} {\bibinfo {author} {\bibfnamefont {J.~E.}\ \bibnamefont
  {Hoffman}}, \bibinfo {author} {\bibfnamefont {E.~W.}\ \bibnamefont {Hudson}},
  \bibinfo {author} {\bibfnamefont {K.~M.}\ \bibnamefont {Lang}}, \bibinfo
  {author} {\bibfnamefont {V.}~\bibnamefont {Madhavan}}, \bibinfo {author}
  {\bibfnamefont {H.}~\bibnamefont {Eisaki}}, \bibinfo {author} {\bibfnamefont
  {S.}~\bibnamefont {Uchida}}, \ and\ \bibinfo {author} {\bibfnamefont {J.~C.}\
  \bibnamefont {Davis}},\ }\href {\doibase 10.1126/science.1066974} {\bibfield
  {journal} {\bibinfo  {journal} {Science}\ }\textbf {\bibinfo {volume}
  {295}},\ \bibinfo {pages} {466} (\bibinfo {year} {2002})}\BibitemShut
  {NoStop}%
\bibitem [{\citenamefont {Fischer}\ \emph {et~al.}(2007)\citenamefont
  {Fischer}, \citenamefont {Kugler}, \citenamefont {Maggio-Aprile},
  \citenamefont {Berthod},\ and\ \citenamefont
  {Renner}}]{fischer_scanning_2007}%
  \BibitemOpen
  \bibfield  {author} {\bibinfo {author} {\bibfnamefont {O.~y.}\ \bibnamefont
  {Fischer}}, \bibinfo {author} {\bibfnamefont {M.}~\bibnamefont {Kugler}},
  \bibinfo {author} {\bibfnamefont {I.}~\bibnamefont {Maggio-Aprile}}, \bibinfo
  {author} {\bibfnamefont {C.}~\bibnamefont {Berthod}}, \ and\ \bibinfo
  {author} {\bibfnamefont {C.}~\bibnamefont {Renner}},\ }\href {\doibase
  10.1103/RevModPhys.79.353} {\bibfield  {journal} {\bibinfo  {journal}
  {Reviews of Modern Physics}\ }\textbf {\bibinfo {volume} {79}},\ \bibinfo
  {pages} {353} (\bibinfo {year} {2007})}\BibitemShut {NoStop}%
\bibitem [{\citenamefont {Maggio-Aprile}\ \emph {et~al.}(1995)\citenamefont
  {Maggio-Aprile}, \citenamefont {Renner}, \citenamefont {Erb}, \citenamefont
  {Walker},\ and\ \citenamefont {Fischer}}]{maggio-aprile_direct_1995}%
  \BibitemOpen
  \bibfield  {author} {\bibinfo {author} {\bibfnamefont {I.}~\bibnamefont
  {Maggio-Aprile}}, \bibinfo {author} {\bibfnamefont {C.}~\bibnamefont
  {Renner}}, \bibinfo {author} {\bibfnamefont {A.}~\bibnamefont {Erb}},
  \bibinfo {author} {\bibfnamefont {E.}~\bibnamefont {Walker}}, \ and\ \bibinfo
  {author} {\bibfnamefont {O.~.}\ \bibnamefont {Fischer}},\ }\href {\doibase
  10.1103/PhysRevLett.75.2754} {\bibfield  {journal} {\bibinfo  {journal}
  {Physical Review Letters}\ }\textbf {\bibinfo {volume} {75}},\ \bibinfo
  {pages} {2754} (\bibinfo {year} {1995})}\BibitemShut {NoStop}%
\bibitem [{\citenamefont {Pan}\ \emph {et~al.}(2000)\citenamefont {Pan},
  \citenamefont {Hudson}, \citenamefont {Gupta}, \citenamefont {Ng},
  \citenamefont {Eisaki}, \citenamefont {Uchida},\ and\ \citenamefont
  {Davis}}]{pan_stm_2000}%
  \BibitemOpen
  \bibfield  {author} {\bibinfo {author} {\bibfnamefont {S.~H.}\ \bibnamefont
  {Pan}}, \bibinfo {author} {\bibfnamefont {E.~W.}\ \bibnamefont {Hudson}},
  \bibinfo {author} {\bibfnamefont {A.~K.}\ \bibnamefont {Gupta}}, \bibinfo
  {author} {\bibfnamefont {K.-W.}\ \bibnamefont {Ng}}, \bibinfo {author}
  {\bibfnamefont {H.}~\bibnamefont {Eisaki}}, \bibinfo {author} {\bibfnamefont
  {S.}~\bibnamefont {Uchida}}, \ and\ \bibinfo {author} {\bibfnamefont {J.~C.}\
  \bibnamefont {Davis}},\ }\href {\doibase 10.1103/PhysRevLett.85.1536}
  {\bibfield  {journal} {\bibinfo  {journal} {Physical Review Letters}\
  }\textbf {\bibinfo {volume} {85}},\ \bibinfo {pages} {1536} (\bibinfo {year}
  {2000})}\BibitemShut {NoStop}%
\bibitem [{\citenamefont {Hoogenboom}\ \emph {et~al.}(2000)\citenamefont
  {Hoogenboom}, \citenamefont {Renner}, \citenamefont {Revaz}, \citenamefont
  {Maggio-Aprile},\ and\ \citenamefont {Fischer}}]{hoogenboom_low-energy_2000}%
  \BibitemOpen
  \bibfield  {author} {\bibinfo {author} {\bibfnamefont {B.~W.}\ \bibnamefont
  {Hoogenboom}}, \bibinfo {author} {\bibfnamefont {C.}~\bibnamefont {Renner}},
  \bibinfo {author} {\bibfnamefont {B.}~\bibnamefont {Revaz}}, \bibinfo
  {author} {\bibfnamefont {I.}~\bibnamefont {Maggio-Aprile}}, \ and\ \bibinfo
  {author} {\bibfnamefont {O.}~\bibnamefont {Fischer}},\ }\href {\doibase
  10.1016/S0921-4534(99)00720-0} {\bibfield  {journal} {\bibinfo  {journal}
  {Physica C: Superconductivity}\ }\textbf {\bibinfo {volume} {332}},\ \bibinfo
  {pages} {440} (\bibinfo {year} {2000})}\BibitemShut {NoStop}%
\bibitem [{\citenamefont {Levy}\ \emph {et~al.}(2005)\citenamefont {Levy},
  \citenamefont {Kugler}, \citenamefont {Manuel}, \citenamefont {Fischer},\
  and\ \citenamefont {Li}}]{levy_fourfold_2005}%
  \BibitemOpen
  \bibfield  {author} {\bibinfo {author} {\bibfnamefont {G.}~\bibnamefont
  {Levy}}, \bibinfo {author} {\bibfnamefont {M.}~\bibnamefont {Kugler}},
  \bibinfo {author} {\bibfnamefont {A.~A.}\ \bibnamefont {Manuel}}, \bibinfo
  {author} {\bibfnamefont {O.~y.}\ \bibnamefont {Fischer}}, \ and\ \bibinfo
  {author} {\bibfnamefont {M.}~\bibnamefont {Li}},\ }\href {\doibase
  10.1103/PhysRevLett.95.257005} {\bibfield  {journal} {\bibinfo  {journal}
  {Physical Review Letters}\ }\textbf {\bibinfo {volume} {95}},\ \bibinfo
  {pages} {257005} (\bibinfo {year} {2005})}\BibitemShut {NoStop}%
\bibitem [{\citenamefont {Bruér}\ \emph {et~al.}(2016)\citenamefont {Bruér},
  \citenamefont {Maggio-Aprile}, \citenamefont {Jenkins}, \citenamefont
  {Ristić}, \citenamefont {Erb}, \citenamefont {Berthod}, \citenamefont
  {Fischer},\ and\ \citenamefont {Renner}}]{bruer_revisiting_2016}%
  \BibitemOpen
  \bibfield  {author} {\bibinfo {author} {\bibfnamefont {J.}~\bibnamefont
  {Bruér}}, \bibinfo {author} {\bibfnamefont {I.}~\bibnamefont
  {Maggio-Aprile}}, \bibinfo {author} {\bibfnamefont {N.}~\bibnamefont
  {Jenkins}}, \bibinfo {author} {\bibfnamefont {Z.}~\bibnamefont {Ristić}},
  \bibinfo {author} {\bibfnamefont {A.}~\bibnamefont {Erb}}, \bibinfo {author}
  {\bibfnamefont {C.}~\bibnamefont {Berthod}}, \bibinfo {author} {\bibfnamefont
  {O.~y.}\ \bibnamefont {Fischer}}, \ and\ \bibinfo {author} {\bibfnamefont
  {C.}~\bibnamefont {Renner}},\ }\href {\doibase 10.1038/ncomms11139}
  {\bibfield  {journal} {\bibinfo  {journal} {Nature Communications}\ }\textbf
  {\bibinfo {volume} {7}},\ \bibinfo {pages} {11139} (\bibinfo {year}
  {2016})},\ \bibinfo {note} {arXiv: 1507.06775}\BibitemShut {NoStop}%
\bibitem [{\citenamefont {McElroy}\ \emph {et~al.}(2005)\citenamefont
  {McElroy}, \citenamefont {Lee}, \citenamefont {Hoffman}, \citenamefont
  {Lang}, \citenamefont {Lee}, \citenamefont {Hudson}, \citenamefont {Eisaki},
  \citenamefont {Uchida},\ and\ \citenamefont
  {Davis}}]{mcelroy_coincidence_2005}%
  \BibitemOpen
  \bibfield  {author} {\bibinfo {author} {\bibfnamefont {K.}~\bibnamefont
  {McElroy}}, \bibinfo {author} {\bibfnamefont {D.-H.}\ \bibnamefont {Lee}},
  \bibinfo {author} {\bibfnamefont {J.~E.}\ \bibnamefont {Hoffman}}, \bibinfo
  {author} {\bibfnamefont {K.~M.}\ \bibnamefont {Lang}}, \bibinfo {author}
  {\bibfnamefont {J.}~\bibnamefont {Lee}}, \bibinfo {author} {\bibfnamefont
  {E.~W.}\ \bibnamefont {Hudson}}, \bibinfo {author} {\bibfnamefont
  {H.}~\bibnamefont {Eisaki}}, \bibinfo {author} {\bibfnamefont
  {S.}~\bibnamefont {Uchida}}, \ and\ \bibinfo {author} {\bibfnamefont {J.~C.}\
  \bibnamefont {Davis}},\ }\href {\doibase 10.1103/PhysRevLett.94.197005}
  {\bibfield  {journal} {\bibinfo  {journal} {Physical Review Letters}\
  }\textbf {\bibinfo {volume} {94}},\ \bibinfo {pages} {197005} (\bibinfo
  {year} {2005})}\BibitemShut {NoStop}%
\bibitem [{\citenamefont {Kohsaka}\ \emph {et~al.}(2007)\citenamefont
  {Kohsaka}, \citenamefont {Taylor}, \citenamefont {Fujita}, \citenamefont
  {Schmidt}, \citenamefont {Lupien}, \citenamefont {Hanaguri}, \citenamefont
  {Azuma}, \citenamefont {Takano}, \citenamefont {Eisaki}, \citenamefont
  {Takagi}, \citenamefont {Uchida},\ and\ \citenamefont
  {Davis}}]{kohsaka_intrinsic_2007}%
  \BibitemOpen
  \bibfield  {author} {\bibinfo {author} {\bibfnamefont {Y.}~\bibnamefont
  {Kohsaka}}, \bibinfo {author} {\bibfnamefont {C.}~\bibnamefont {Taylor}},
  \bibinfo {author} {\bibfnamefont {K.}~\bibnamefont {Fujita}}, \bibinfo
  {author} {\bibfnamefont {A.}~\bibnamefont {Schmidt}}, \bibinfo {author}
  {\bibfnamefont {C.}~\bibnamefont {Lupien}}, \bibinfo {author} {\bibfnamefont
  {T.}~\bibnamefont {Hanaguri}}, \bibinfo {author} {\bibfnamefont
  {M.}~\bibnamefont {Azuma}}, \bibinfo {author} {\bibfnamefont
  {M.}~\bibnamefont {Takano}}, \bibinfo {author} {\bibfnamefont
  {H.}~\bibnamefont {Eisaki}}, \bibinfo {author} {\bibfnamefont
  {H.}~\bibnamefont {Takagi}}, \bibinfo {author} {\bibfnamefont
  {S.}~\bibnamefont {Uchida}}, \ and\ \bibinfo {author} {\bibfnamefont {J.~C.}\
  \bibnamefont {Davis}},\ }\href {\doibase 10.1126/science.1138584} {\bibfield
  {journal} {\bibinfo  {journal} {Science}\ }\textbf {\bibinfo {volume}
  {315}},\ \bibinfo {pages} {1380} (\bibinfo {year} {2007})}\BibitemShut
  {NoStop}%
\bibitem [{\citenamefont {McElroy}\ \emph {et~al.}(2004)\citenamefont
  {McElroy}, \citenamefont {Lee}, \citenamefont {Hoffman}, \citenamefont
  {Lang}, \citenamefont {Hudson}, \citenamefont {Eisaki}, \citenamefont
  {Uchida}, \citenamefont {Lee},\ and\ \citenamefont
  {Davis}}]{mcelroy_homogenous_2004}%
  \BibitemOpen
  \bibfield  {author} {\bibinfo {author} {\bibfnamefont {K.}~\bibnamefont
  {McElroy}}, \bibinfo {author} {\bibfnamefont {D.-H.}\ \bibnamefont {Lee}},
  \bibinfo {author} {\bibfnamefont {J.~E.}\ \bibnamefont {Hoffman}}, \bibinfo
  {author} {\bibfnamefont {K.~M.}\ \bibnamefont {Lang}}, \bibinfo {author}
  {\bibfnamefont {E.~W.}\ \bibnamefont {Hudson}}, \bibinfo {author}
  {\bibfnamefont {H.}~\bibnamefont {Eisaki}}, \bibinfo {author} {\bibfnamefont
  {S.}~\bibnamefont {Uchida}}, \bibinfo {author} {\bibfnamefont
  {J.}~\bibnamefont {Lee}}, \ and\ \bibinfo {author} {\bibfnamefont {J.~C.}\
  \bibnamefont {Davis}},\ }\href {http://arxiv.org/abs/cond-mat/0404005}
  {\bibfield  {journal} {\bibinfo  {journal} {arXiv:cond-mat/0404005}\ }
  (\bibinfo {year} {2004})},\ \bibinfo {note} {arXiv:
  cond-mat/0404005}\BibitemShut {NoStop}%
\bibitem [{\citenamefont {Andersen}\ \emph {et~al.}(2003)\citenamefont
  {Andersen}, \citenamefont {Hedegård},\ and\ \citenamefont
  {Bruus}}]{andersen_checkerboard_2003}%
  \BibitemOpen
  \bibfield  {author} {\bibinfo {author} {\bibfnamefont {B.~M.~l.}\
  \bibnamefont {Andersen}}, \bibinfo {author} {\bibfnamefont {P.}~\bibnamefont
  {Hedegård}}, \ and\ \bibinfo {author} {\bibfnamefont {H.}~\bibnamefont
  {Bruus}},\ }\href {\doibase 10.1103/PhysRevB.67.134528} {\bibfield  {journal}
  {\bibinfo  {journal} {Physical Review B}\ }\textbf {\bibinfo {volume} {67}},\
  \bibinfo {pages} {134528} (\bibinfo {year} {2003})}\BibitemShut {NoStop}%
\bibitem [{\citenamefont {Kishine}\ \emph {et~al.}(2001)\citenamefont
  {Kishine}, \citenamefont {Lee},\ and\ \citenamefont
  {Wen}}]{kishine_staggered_2001}%
  \BibitemOpen
  \bibfield  {author} {\bibinfo {author} {\bibfnamefont {J.-i.}\ \bibnamefont
  {Kishine}}, \bibinfo {author} {\bibfnamefont {P.~A.}\ \bibnamefont {Lee}}, \
  and\ \bibinfo {author} {\bibfnamefont {X.-G.}\ \bibnamefont {Wen}},\ }\href
  {\doibase 10.1103/PhysRevLett.86.5365} {\bibfield  {journal} {\bibinfo
  {journal} {Physical Review Letters}\ }\textbf {\bibinfo {volume} {86}},\
  \bibinfo {pages} {5365} (\bibinfo {year} {2001})}\BibitemShut {NoStop}%
\bibitem [{\citenamefont {Lee}\ and\ \citenamefont
  {Wen}(2001)}]{lee_vortex_2001}%
  \BibitemOpen
  \bibfield  {author} {\bibinfo {author} {\bibfnamefont {P.~A.}\ \bibnamefont
  {Lee}}\ and\ \bibinfo {author} {\bibfnamefont {X.-G.}\ \bibnamefont {Wen}},\
  }\href {\doibase 10.1103/PhysRevB.63.224517} {\bibfield  {journal} {\bibinfo
  {journal} {Physical Review B}\ }\textbf {\bibinfo {volume} {63}},\ \bibinfo
  {pages} {224517} (\bibinfo {year} {2001})}\BibitemShut {NoStop}%
\bibitem [{\citenamefont {Chen}\ \emph {et~al.}(2002)\citenamefont {Chen},
  \citenamefont {Hu}, \citenamefont {Capponi}, \citenamefont {Arrigoni},\ and\
  \citenamefont {Zhang}}]{chen_antiferromagnetism_2002}%
  \BibitemOpen
  \bibfield  {author} {\bibinfo {author} {\bibfnamefont {H.-D.}\ \bibnamefont
  {Chen}}, \bibinfo {author} {\bibfnamefont {J.-P.}\ \bibnamefont {Hu}},
  \bibinfo {author} {\bibfnamefont {S.}~\bibnamefont {Capponi}}, \bibinfo
  {author} {\bibfnamefont {E.}~\bibnamefont {Arrigoni}}, \ and\ \bibinfo
  {author} {\bibfnamefont {S.-C.}\ \bibnamefont {Zhang}},\ }\href {\doibase
  10.1103/PhysRevLett.89.137004} {\bibfield  {journal} {\bibinfo  {journal}
  {Physical Review Letters}\ }\textbf {\bibinfo {volume} {89}},\ \bibinfo
  {pages} {137004} (\bibinfo {year} {2002})}\BibitemShut {NoStop}%
\bibitem [{\citenamefont {Chen}\ \emph {et~al.}(2004)\citenamefont {Chen},
  \citenamefont {Vafek}, \citenamefont {Yazdani},\ and\ \citenamefont
  {Zhang}}]{chen_pair_2004}%
  \BibitemOpen
  \bibfield  {author} {\bibinfo {author} {\bibfnamefont {H.-D.}\ \bibnamefont
  {Chen}}, \bibinfo {author} {\bibfnamefont {O.}~\bibnamefont {Vafek}},
  \bibinfo {author} {\bibfnamefont {A.}~\bibnamefont {Yazdani}}, \ and\
  \bibinfo {author} {\bibfnamefont {S.-C.}\ \bibnamefont {Zhang}},\ }\href
  {\doibase 10.1103/PhysRevLett.93.187002} {\bibfield  {journal} {\bibinfo
  {journal} {Physical Review Letters}\ }\textbf {\bibinfo {volume} {93}},\
  \bibinfo {pages} {187002} (\bibinfo {year} {2004})}\BibitemShut {NoStop}%
\bibitem [{\citenamefont {Seo}\ \emph {et~al.}(2007)\citenamefont {Seo},
  \citenamefont {Chen},\ and\ \citenamefont {Hu}}]{seo_$d$-wave_2007}%
  \BibitemOpen
  \bibfield  {author} {\bibinfo {author} {\bibfnamefont {K.}~\bibnamefont
  {Seo}}, \bibinfo {author} {\bibfnamefont {H.-D.}\ \bibnamefont {Chen}}, \
  and\ \bibinfo {author} {\bibfnamefont {J.}~\bibnamefont {Hu}},\ }\href
  {\doibase 10.1103/PhysRevB.76.020511} {\bibfield  {journal} {\bibinfo
  {journal} {Physical Review B}\ }\textbf {\bibinfo {volume} {76}},\ \bibinfo
  {pages} {020511} (\bibinfo {year} {2007})}\BibitemShut {NoStop}%
\bibitem [{\citenamefont {Seo}\ \emph {et~al.}(2008)\citenamefont {Seo},
  \citenamefont {Chen},\ and\ \citenamefont {Hu}}]{seo_complementary_2008}%
  \BibitemOpen
  \bibfield  {author} {\bibinfo {author} {\bibfnamefont {K.}~\bibnamefont
  {Seo}}, \bibinfo {author} {\bibfnamefont {H.-D.}\ \bibnamefont {Chen}}, \
  and\ \bibinfo {author} {\bibfnamefont {J.}~\bibnamefont {Hu}},\ }\href
  {\doibase 10.1103/PhysRevB.78.094510} {\bibfield  {journal} {\bibinfo
  {journal} {Physical Review B}\ }\textbf {\bibinfo {volume} {78}},\ \bibinfo
  {pages} {094510} (\bibinfo {year} {2008})}\BibitemShut {NoStop}%
\bibitem [{\citenamefont {Agterberg}\ and\ \citenamefont
  {Garaud}(2015)}]{agterberg_checkerboard_2015}%
  \BibitemOpen
  \bibfield  {author} {\bibinfo {author} {\bibfnamefont {D.~F.}\ \bibnamefont
  {Agterberg}}\ and\ \bibinfo {author} {\bibfnamefont {J.}~\bibnamefont
  {Garaud}},\ }\href {\doibase 10.1103/PhysRevB.91.104512} {\bibfield
  {journal} {\bibinfo  {journal} {Physical Review B}\ }\textbf {\bibinfo
  {volume} {91}},\ \bibinfo {pages} {104512} (\bibinfo {year}
  {2015})}\BibitemShut {NoStop}%
\bibitem [{\citenamefont {Lee}(2014)}]{lee_amperean_2014}%
  \BibitemOpen
  \bibfield  {author} {\bibinfo {author} {\bibfnamefont {P.~A.}\ \bibnamefont
  {Lee}},\ }\href {\doibase 10.1103/PhysRevX.4.031017} {\bibfield  {journal}
  {\bibinfo  {journal} {Physical Review X}\ }\textbf {\bibinfo {volume} {4}},\
  \bibinfo {pages} {031017} (\bibinfo {year} {2014})}\BibitemShut {NoStop}%
\bibitem [{\citenamefont {Himeda}\ \emph {et~al.}(2002)\citenamefont {Himeda},
  \citenamefont {Kato},\ and\ \citenamefont {Ogata}}]{himeda_stripe_2002}%
  \BibitemOpen
  \bibfield  {author} {\bibinfo {author} {\bibfnamefont {A.}~\bibnamefont
  {Himeda}}, \bibinfo {author} {\bibfnamefont {T.}~\bibnamefont {Kato}}, \ and\
  \bibinfo {author} {\bibfnamefont {M.}~\bibnamefont {Ogata}},\ }\href
  {\doibase 10.1103/PhysRevLett.88.117001} {\bibfield  {journal} {\bibinfo
  {journal} {Physical Review Letters}\ }\textbf {\bibinfo {volume} {88}},\
  \bibinfo {pages} {117001} (\bibinfo {year} {2002})}\BibitemShut {NoStop}%
\bibitem [{\citenamefont {Raczkowski}\ \emph {et~al.}(2007)\citenamefont
  {Raczkowski}, \citenamefont {Capello}, \citenamefont {Poilblanc},
  \citenamefont {Frésard},\ and\ \citenamefont
  {Oleś}}]{raczkowski_unidirectional_2007}%
  \BibitemOpen
  \bibfield  {author} {\bibinfo {author} {\bibfnamefont {M.}~\bibnamefont
  {Raczkowski}}, \bibinfo {author} {\bibfnamefont {M.}~\bibnamefont {Capello}},
  \bibinfo {author} {\bibfnamefont {D.}~\bibnamefont {Poilblanc}}, \bibinfo
  {author} {\bibfnamefont {R.}~\bibnamefont {Frésard}}, \ and\ \bibinfo
  {author} {\bibfnamefont {A.~M.}\ \bibnamefont {Oleś}},\ }\href {\doibase
  10.1103/PhysRevB.76.140505} {\bibfield  {journal} {\bibinfo  {journal}
  {Physical Review B}\ }\textbf {\bibinfo {volume} {76}},\ \bibinfo {pages}
  {140505} (\bibinfo {year} {2007})}\BibitemShut {NoStop}%
\bibitem [{\citenamefont {Wang}\ \emph {et~al.}(2015)\citenamefont {Wang},
  \citenamefont {Agterberg},\ and\ \citenamefont
  {Chubukov}}]{wang_coexistence_2015}%
  \BibitemOpen
  \bibfield  {author} {\bibinfo {author} {\bibfnamefont {Y.}~\bibnamefont
  {Wang}}, \bibinfo {author} {\bibfnamefont {D.~F.}\ \bibnamefont {Agterberg}},
  \ and\ \bibinfo {author} {\bibfnamefont {A.}~\bibnamefont {Chubukov}},\
  }\href {\doibase 10.1103/PhysRevLett.114.197001} {\bibfield  {journal}
  {\bibinfo  {journal} {Physical Review Letters}\ }\textbf {\bibinfo {volume}
  {114}},\ \bibinfo {pages} {197001} (\bibinfo {year} {2015})}\BibitemShut
  {NoStop}%
\bibitem [{\citenamefont {Freire}\ \emph {et~al.}(2015)\citenamefont {Freire},
  \citenamefont {de~Carvalho},\ and\ \citenamefont
  {Pépin}}]{freire_renormalization_2015}%
  \BibitemOpen
  \bibfield  {author} {\bibinfo {author} {\bibfnamefont {H.}~\bibnamefont
  {Freire}}, \bibinfo {author} {\bibfnamefont {V.~S.}\ \bibnamefont
  {de~Carvalho}}, \ and\ \bibinfo {author} {\bibfnamefont {C.}~\bibnamefont
  {Pépin}},\ }\href {\doibase 10.1103/PhysRevB.92.045132} {\bibfield
  {journal} {\bibinfo  {journal} {Physical Review B}\ }\textbf {\bibinfo
  {volume} {92}},\ \bibinfo {pages} {045132} (\bibinfo {year}
  {2015})}\BibitemShut {NoStop}%
\bibitem [{\citenamefont {Fradkin}\ \emph {et~al.}(2015)\citenamefont
  {Fradkin}, \citenamefont {Kivelson},\ and\ \citenamefont
  {Tranquada}}]{fradkin_colloquium_2015}%
  \BibitemOpen
  \bibfield  {author} {\bibinfo {author} {\bibfnamefont {E.}~\bibnamefont
  {Fradkin}}, \bibinfo {author} {\bibfnamefont {S.~A.}\ \bibnamefont
  {Kivelson}}, \ and\ \bibinfo {author} {\bibfnamefont {J.~M.}\ \bibnamefont
  {Tranquada}},\ }\href {\doibase 10.1103/RevModPhys.87.457} {\bibfield
  {journal} {\bibinfo  {journal} {Reviews of Modern Physics}\ }\textbf
  {\bibinfo {volume} {87}},\ \bibinfo {pages} {457} (\bibinfo {year}
  {2015})}\BibitemShut {NoStop}%
\bibitem [{\citenamefont {Hamidian}\ \emph
  {et~al.}(2016{\natexlab{b}})\citenamefont {Hamidian}, \citenamefont {Edkins},
  \citenamefont {Joo}, \citenamefont {Kostin}, \citenamefont {Eisaki},
  \citenamefont {Uchida}, \citenamefont {Lawler}, \citenamefont {Kim},
  \citenamefont {Mackenzie}, \citenamefont {Fujita}, \citenamefont {Lee},\ and\
  \citenamefont {Davis}}]{hamidian_detection_2016}%
  \BibitemOpen
  \bibfield  {author} {\bibinfo {author} {\bibfnamefont {M.~H.}\ \bibnamefont
  {Hamidian}}, \bibinfo {author} {\bibfnamefont {S.~D.}\ \bibnamefont
  {Edkins}}, \bibinfo {author} {\bibfnamefont {S.~H.}\ \bibnamefont {Joo}},
  \bibinfo {author} {\bibfnamefont {A.}~\bibnamefont {Kostin}}, \bibinfo
  {author} {\bibfnamefont {H.}~\bibnamefont {Eisaki}}, \bibinfo {author}
  {\bibfnamefont {S.}~\bibnamefont {Uchida}}, \bibinfo {author} {\bibfnamefont
  {M.~J.}\ \bibnamefont {Lawler}}, \bibinfo {author} {\bibfnamefont {E.-A.}\
  \bibnamefont {Kim}}, \bibinfo {author} {\bibfnamefont {A.~P.}\ \bibnamefont
  {Mackenzie}}, \bibinfo {author} {\bibfnamefont {K.}~\bibnamefont {Fujita}},
  \bibinfo {author} {\bibfnamefont {J.}~\bibnamefont {Lee}}, \ and\ \bibinfo
  {author} {\bibfnamefont {J.~C.~S.}\ \bibnamefont {Davis}},\ }\href {\doibase
  10.1038/nature17411} {\bibfield  {journal} {\bibinfo  {journal} {Nature}\
  }\textbf {\bibinfo {volume} {532}},\ \bibinfo {pages} {343} (\bibinfo {year}
  {2016}{\natexlab{b}})}\BibitemShut {NoStop}%
\bibitem [{\citenamefont {Fujita}\ \emph {et~al.}(2014)\citenamefont {Fujita},
  \citenamefont {Hamidian}, \citenamefont {Edkins}, \citenamefont {Kim},
  \citenamefont {Kohsaka}, \citenamefont {Azuma}, \citenamefont {Takano},
  \citenamefont {Takagi}, \citenamefont {Eisaki}, \citenamefont {Uchida},
  \citenamefont {Allais}, \citenamefont {Lawler}, \citenamefont {Kim},
  \citenamefont {Sachdev},\ and\ \citenamefont {Davis}}]{fujita_direct_2014}%
  \BibitemOpen
  \bibfield  {author} {\bibinfo {author} {\bibfnamefont {K.}~\bibnamefont
  {Fujita}}, \bibinfo {author} {\bibfnamefont {M.~H.}\ \bibnamefont
  {Hamidian}}, \bibinfo {author} {\bibfnamefont {S.~D.}\ \bibnamefont
  {Edkins}}, \bibinfo {author} {\bibfnamefont {C.~K.}\ \bibnamefont {Kim}},
  \bibinfo {author} {\bibfnamefont {Y.}~\bibnamefont {Kohsaka}}, \bibinfo
  {author} {\bibfnamefont {M.}~\bibnamefont {Azuma}}, \bibinfo {author}
  {\bibfnamefont {M.}~\bibnamefont {Takano}}, \bibinfo {author} {\bibfnamefont
  {H.}~\bibnamefont {Takagi}}, \bibinfo {author} {\bibfnamefont
  {H.}~\bibnamefont {Eisaki}}, \bibinfo {author} {\bibfnamefont {S.-i.}\
  \bibnamefont {Uchida}}, \bibinfo {author} {\bibfnamefont {A.}~\bibnamefont
  {Allais}}, \bibinfo {author} {\bibfnamefont {M.~J.}\ \bibnamefont {Lawler}},
  \bibinfo {author} {\bibfnamefont {E.-A.}\ \bibnamefont {Kim}}, \bibinfo
  {author} {\bibfnamefont {S.}~\bibnamefont {Sachdev}}, \ and\ \bibinfo
  {author} {\bibfnamefont {J.~C.~S.}\ \bibnamefont {Davis}},\ }\href {\doibase
  10.1073/pnas.1406297111} {\bibfield  {journal} {\bibinfo  {journal}
  {Proceedings of the National Academy of Sciences}\ }\textbf {\bibinfo
  {volume} {111}},\ \bibinfo {pages} {E3026} (\bibinfo {year}
  {2014})}\BibitemShut {NoStop}%
\bibitem [{\citenamefont {Machida}\ \emph {et~al.}(2016)\citenamefont
  {Machida}, \citenamefont {Kohsaka}, \citenamefont {Matsuoka}, \citenamefont
  {Iwaya}, \citenamefont {Hanaguri},\ and\ \citenamefont
  {Tamegai}}]{machida_bipartite_2016}%
  \BibitemOpen
  \bibfield  {author} {\bibinfo {author} {\bibfnamefont {T.}~\bibnamefont
  {Machida}}, \bibinfo {author} {\bibfnamefont {Y.}~\bibnamefont {Kohsaka}},
  \bibinfo {author} {\bibfnamefont {K.}~\bibnamefont {Matsuoka}}, \bibinfo
  {author} {\bibfnamefont {K.}~\bibnamefont {Iwaya}}, \bibinfo {author}
  {\bibfnamefont {T.}~\bibnamefont {Hanaguri}}, \ and\ \bibinfo {author}
  {\bibfnamefont {T.}~\bibnamefont {Tamegai}},\ }\href {\doibase
  10.1038/ncomms11747} {\bibfield  {journal} {\bibinfo  {journal} {Nature
  Communications}\ }\textbf {\bibinfo {volume} {7}},\ \bibinfo {pages} {11747}
  (\bibinfo {year} {2016})}\BibitemShut {NoStop}%
\bibitem [{\citenamefont {Allais}\ \emph {et~al.}(2014)\citenamefont {Allais},
  \citenamefont {Chowdhury},\ and\ \citenamefont
  {Sachdev}}]{allais_connecting_2014}%
  \BibitemOpen
  \bibfield  {author} {\bibinfo {author} {\bibfnamefont {A.}~\bibnamefont
  {Allais}}, \bibinfo {author} {\bibfnamefont {D.}~\bibnamefont {Chowdhury}}, \
  and\ \bibinfo {author} {\bibfnamefont {S.}~\bibnamefont {Sachdev}},\ }\href
  {\doibase 10.1038/ncomms6771} {\bibfield  {journal} {\bibinfo  {journal}
  {Nature Communications}\ }\textbf {\bibinfo {volume} {5}} (\bibinfo {year}
  {2014}),\ 10.1038/ncomms6771}\BibitemShut {NoStop}%
\bibitem [{\citenamefont {Yang}\ \emph {et~al.}(2006)\citenamefont {Yang},
  \citenamefont {Rice},\ and\ \citenamefont
  {Zhang}}]{yang_phenomenological_2006}%
  \BibitemOpen
  \bibfield  {author} {\bibinfo {author} {\bibfnamefont {K.-Y.}\ \bibnamefont
  {Yang}}, \bibinfo {author} {\bibfnamefont {T.~M.}\ \bibnamefont {Rice}}, \
  and\ \bibinfo {author} {\bibfnamefont {F.-C.}\ \bibnamefont {Zhang}},\ }\href
  {\doibase 10.1103/PhysRevB.73.174501} {\bibfield  {journal} {\bibinfo
  {journal} {Physical Review B}\ }\textbf {\bibinfo {volume} {73}},\ \bibinfo
  {pages} {174501} (\bibinfo {year} {2006})}\BibitemShut {NoStop}%
\bibitem [{\citenamefont {Qi}\ and\ \citenamefont
  {Sachdev}(2010)}]{qi_effective_2010}%
  \BibitemOpen
  \bibfield  {author} {\bibinfo {author} {\bibfnamefont {Y.}~\bibnamefont
  {Qi}}\ and\ \bibinfo {author} {\bibfnamefont {S.}~\bibnamefont {Sachdev}},\
  }\href {\doibase 10.1103/PhysRevB.81.115129} {\bibfield  {journal} {\bibinfo
  {journal} {Physical Review B}\ }\textbf {\bibinfo {volume} {81}},\ \bibinfo
  {pages} {115129} (\bibinfo {year} {2010})}\BibitemShut {NoStop}%
\bibitem [{\citenamefont {Vishik}\ \emph {et~al.}(2010)\citenamefont {Vishik},
  \citenamefont {Lee}, \citenamefont {He}, \citenamefont {Hashimoto},
  \citenamefont {Hussain}, \citenamefont {Devereaux},\ and\ \citenamefont
  {Shen}}]{vishik_arpes_2010}%
  \BibitemOpen
  \bibfield  {author} {\bibinfo {author} {\bibfnamefont {I.~M.}\ \bibnamefont
  {Vishik}}, \bibinfo {author} {\bibfnamefont {W.~S.}\ \bibnamefont {Lee}},
  \bibinfo {author} {\bibfnamefont {R.-H.}\ \bibnamefont {He}}, \bibinfo
  {author} {\bibfnamefont {M.}~\bibnamefont {Hashimoto}}, \bibinfo {author}
  {\bibfnamefont {Z.}~\bibnamefont {Hussain}}, \bibinfo {author} {\bibfnamefont
  {T.~P.}\ \bibnamefont {Devereaux}}, \ and\ \bibinfo {author} {\bibfnamefont
  {Z.-X.}\ \bibnamefont {Shen}},\ }\href {\doibase
  10.1088/1367-2630/12/10/105008} {\bibfield  {journal} {\bibinfo  {journal}
  {New Journal of Physics}\ }\textbf {\bibinfo {volume} {12}},\ \bibinfo
  {pages} {105008} (\bibinfo {year} {2010})}\BibitemShut {NoStop}%
\bibitem [{\citenamefont {Sachdev}\ and\ \citenamefont
  {La~Placa}(2013)}]{sachdev_bond_2013}%
  \BibitemOpen
  \bibfield  {author} {\bibinfo {author} {\bibfnamefont {S.}~\bibnamefont
  {Sachdev}}\ and\ \bibinfo {author} {\bibfnamefont {R.}~\bibnamefont
  {La~Placa}},\ }\href {\doibase 10.1103/PhysRevLett.111.027202} {\bibfield
  {journal} {\bibinfo  {journal} {Physical Review Letters}\ }\textbf {\bibinfo
  {volume} {111}},\ \bibinfo {pages} {027202} (\bibinfo {year}
  {2013})}\BibitemShut {NoStop}%
\bibitem [{\citenamefont {Lee}\ \emph {et~al.}(2016)\citenamefont {Lee},
  \citenamefont {Kivelson},\ and\ \citenamefont {Kim}}]{lee_cold-spots_2016}%
  \BibitemOpen
  \bibfield  {author} {\bibinfo {author} {\bibfnamefont {K.}~\bibnamefont
  {Lee}}, \bibinfo {author} {\bibfnamefont {S.~A.}\ \bibnamefont {Kivelson}}, \
  and\ \bibinfo {author} {\bibfnamefont {E.-A.}\ \bibnamefont {Kim}},\ }\href
  {\doibase 10.1103/PhysRevB.94.014204} {\bibfield  {journal} {\bibinfo
  {journal} {Physical Review B}\ }\textbf {\bibinfo {volume} {94}},\ \bibinfo
  {pages} {014204} (\bibinfo {year} {2016})}\BibitemShut {NoStop}%
\bibitem [{\citenamefont {Andersen}\ and\ \citenamefont
  {Hirschfeld}(2009)}]{andersen_extinction_2009}%
  \BibitemOpen
  \bibfield  {author} {\bibinfo {author} {\bibfnamefont {B.~M.}\ \bibnamefont
  {Andersen}}\ and\ \bibinfo {author} {\bibfnamefont {P.~J.}\ \bibnamefont
  {Hirschfeld}},\ }\href {\doibase 10.1103/PhysRevB.79.144515} {\bibfield
  {journal} {\bibinfo  {journal} {Physical Review B}\ }\textbf {\bibinfo
  {volume} {79}},\ \bibinfo {pages} {144515} (\bibinfo {year}
  {2009})}\BibitemShut {NoStop}%
\bibitem [{\citenamefont {Nordheim}(1931)}]{nordheim_electron_1931}%
  \BibitemOpen
  \bibfield  {author} {\bibinfo {author} {\bibfnamefont {L.}~\bibnamefont
  {Nordheim}},\ }\href@noop {} {\bibfield  {journal} {\bibinfo  {journal} {Ann.
  Phys.}\ }\textbf {\bibinfo {volume} {9}},\ \bibinfo {pages} {607} (\bibinfo
  {year} {1931})}\BibitemShut {NoStop}%
\bibitem [{\citenamefont {Bellaiche}\ and\ \citenamefont
  {Vanderbilt}(2000)}]{bellaiche_virtual_2000}%
  \BibitemOpen
  \bibfield  {author} {\bibinfo {author} {\bibfnamefont {L.}~\bibnamefont
  {Bellaiche}}\ and\ \bibinfo {author} {\bibfnamefont {D.}~\bibnamefont
  {Vanderbilt}},\ }\href {\doibase 10.1103/PhysRevB.61.7877} {\bibfield
  {journal} {\bibinfo  {journal} {Physical Review B}\ }\textbf {\bibinfo
  {volume} {61}},\ \bibinfo {pages} {7877} (\bibinfo {year}
  {2000})}\BibitemShut {NoStop}%
\bibitem [{\citenamefont {Garg}\ \emph {et~al.}(2008)\citenamefont {Garg},
  \citenamefont {Randeria},\ and\ \citenamefont {Trivedi}}]{garg_strong_2008}%
  \BibitemOpen
  \bibfield  {author} {\bibinfo {author} {\bibfnamefont {A.}~\bibnamefont
  {Garg}}, \bibinfo {author} {\bibfnamefont {M.}~\bibnamefont {Randeria}}, \
  and\ \bibinfo {author} {\bibfnamefont {N.}~\bibnamefont {Trivedi}},\ }\href
  {\doibase 10.1038/nphys1026} {\bibfield  {journal} {\bibinfo  {journal}
  {Nature Physics}\ }\textbf {\bibinfo {volume} {4}},\ \bibinfo {pages} {762}
  (\bibinfo {year} {2008})}\BibitemShut {NoStop}%
\bibitem [{\citenamefont {Tang}\ \emph {et~al.}(2016)\citenamefont {Tang},
  \citenamefont {Dobrosavljević},\ and\ \citenamefont
  {Miranda}}]{tang_strong_2016}%
  \BibitemOpen
  \bibfield  {author} {\bibinfo {author} {\bibfnamefont {S.}~\bibnamefont
  {Tang}}, \bibinfo {author} {\bibfnamefont {V.}~\bibnamefont
  {Dobrosavljević}}, \ and\ \bibinfo {author} {\bibfnamefont {E.}~\bibnamefont
  {Miranda}},\ }\href {\doibase 10.1103/PhysRevB.93.195109} {\bibfield
  {journal} {\bibinfo  {journal} {Physical Review B}\ }\textbf {\bibinfo
  {volume} {93}},\ \bibinfo {pages} {195109} (\bibinfo {year}
  {2016})}\BibitemShut {NoStop}%
\bibitem [{\citenamefont {Andersen}\ \emph {et~al.}(1995)\citenamefont
  {Andersen}, \citenamefont {Liechtenstein}, \citenamefont {Jepsen},\ and\
  \citenamefont {Paulsen}}]{andersen_lda_1995}%
  \BibitemOpen
  \bibfield  {author} {\bibinfo {author} {\bibfnamefont {O.}~\bibnamefont
  {Andersen}}, \bibinfo {author} {\bibfnamefont {A.}~\bibnamefont
  {Liechtenstein}}, \bibinfo {author} {\bibfnamefont {O.}~\bibnamefont
  {Jepsen}}, \ and\ \bibinfo {author} {\bibfnamefont {F.}~\bibnamefont
  {Paulsen}},\ }\href {\doibase 10.1016/0022-3697(95)00269-3} {\bibfield
  {journal} {\bibinfo  {journal} {Journal of Physics and Chemistry of Solids}\
  }\textbf {\bibinfo {volume} {56}},\ \bibinfo {pages} {1573} (\bibinfo {year}
  {1995})}\BibitemShut {NoStop}%
\bibitem [{\citenamefont {Liechtenstein}\ \emph {et~al.}(1996)\citenamefont
  {Liechtenstein}, \citenamefont {Gunnarsson}, \citenamefont {Andersen},\ and\
  \citenamefont {Martin}}]{liechtenstein_quasiparticle_1996}%
  \BibitemOpen
  \bibfield  {author} {\bibinfo {author} {\bibfnamefont {A.~I.}\ \bibnamefont
  {Liechtenstein}}, \bibinfo {author} {\bibfnamefont {O.}~\bibnamefont
  {Gunnarsson}}, \bibinfo {author} {\bibfnamefont {O.~K.}\ \bibnamefont
  {Andersen}}, \ and\ \bibinfo {author} {\bibfnamefont {R.~M.}\ \bibnamefont
  {Martin}},\ }\href {\doibase 10.1103/PhysRevB.54.12505} {\bibfield  {journal}
  {\bibinfo  {journal} {Physical Review B}\ }\textbf {\bibinfo {volume} {54}},\
  \bibinfo {pages} {12505} (\bibinfo {year} {1996})}\BibitemShut {NoStop}%
\bibitem [{\citenamefont {Pavarini}\ \emph {et~al.}(2001)\citenamefont
  {Pavarini}, \citenamefont {Dasgupta}, \citenamefont {Saha-Dasgupta},
  \citenamefont {Jepsen},\ and\ \citenamefont
  {Andersen}}]{pavarini_band-structure_2001}%
  \BibitemOpen
  \bibfield  {author} {\bibinfo {author} {\bibfnamefont {E.}~\bibnamefont
  {Pavarini}}, \bibinfo {author} {\bibfnamefont {I.}~\bibnamefont {Dasgupta}},
  \bibinfo {author} {\bibfnamefont {T.}~\bibnamefont {Saha-Dasgupta}}, \bibinfo
  {author} {\bibfnamefont {O.}~\bibnamefont {Jepsen}}, \ and\ \bibinfo {author}
  {\bibfnamefont {O.~K.}\ \bibnamefont {Andersen}},\ }\href {\doibase
  10.1103/PhysRevLett.87.047003} {\bibfield  {journal} {\bibinfo  {journal}
  {Physical Review Letters}\ }\textbf {\bibinfo {volume} {87}} (\bibinfo {year}
  {2001}),\ 10.1103/PhysRevLett.87.047003}\BibitemShut {NoStop}%
\bibitem [{\citenamefont {Markiewicz}\ \emph {et~al.}(2005)\citenamefont
  {Markiewicz}, \citenamefont {Sahrakorpi}, \citenamefont {Lindroos},
  \citenamefont {Lin},\ and\ \citenamefont
  {Bansil}}]{markiewicz_one-band_2005}%
  \BibitemOpen
  \bibfield  {author} {\bibinfo {author} {\bibfnamefont {R.~S.}\ \bibnamefont
  {Markiewicz}}, \bibinfo {author} {\bibfnamefont {S.}~\bibnamefont
  {Sahrakorpi}}, \bibinfo {author} {\bibfnamefont {M.}~\bibnamefont
  {Lindroos}}, \bibinfo {author} {\bibfnamefont {H.}~\bibnamefont {Lin}}, \
  and\ \bibinfo {author} {\bibfnamefont {A.}~\bibnamefont {Bansil}},\ }\href
  {\doibase 10.1103/PhysRevB.72.054519} {\bibfield  {journal} {\bibinfo
  {journal} {Physical Review B}\ }\textbf {\bibinfo {volume} {72}},\ \bibinfo
  {pages} {054519} (\bibinfo {year} {2005})}\BibitemShut {NoStop}%
\bibitem [{\citenamefont {Norman}\ \emph {et~al.}(1995)\citenamefont {Norman},
  \citenamefont {Randeria}, \citenamefont {Ding},\ and\ \citenamefont
  {Campuzano}}]{norman_phenomenological_1995}%
  \BibitemOpen
  \bibfield  {author} {\bibinfo {author} {\bibfnamefont {M.~R.}\ \bibnamefont
  {Norman}}, \bibinfo {author} {\bibfnamefont {M.}~\bibnamefont {Randeria}},
  \bibinfo {author} {\bibfnamefont {H.}~\bibnamefont {Ding}}, \ and\ \bibinfo
  {author} {\bibfnamefont {J.~C.}\ \bibnamefont {Campuzano}},\ }\href {\doibase
  10.1103/PhysRevB.52.615} {\bibfield  {journal} {\bibinfo  {journal} {Physical
  Review B}\ }\textbf {\bibinfo {volume} {52}},\ \bibinfo {pages} {615}
  (\bibinfo {year} {1995})}\BibitemShut {NoStop}%
\bibitem [{\citenamefont {He}\ \emph {et~al.}(2011)\citenamefont {He},
  \citenamefont {Hashimoto}, \citenamefont {Karapetyan}, \citenamefont
  {Koralek}, \citenamefont {Hinton}, \citenamefont {Testaud}, \citenamefont
  {Nathan}, \citenamefont {Yoshida}, \citenamefont {Yao}, \citenamefont
  {Tanaka}, \citenamefont {Meevasana}, \citenamefont {Moore}, \citenamefont
  {Lu}, \citenamefont {Mo}, \citenamefont {Ishikado}, \citenamefont {Eisaki},
  \citenamefont {Hussain}, \citenamefont {Devereaux}, \citenamefont {Kivelson},
  \citenamefont {Orenstein}, \citenamefont {Kapitulnik},\ and\ \citenamefont
  {Shen}}]{he_single-band_2011}%
  \BibitemOpen
  \bibfield  {author} {\bibinfo {author} {\bibfnamefont {R.-H.}\ \bibnamefont
  {He}}, \bibinfo {author} {\bibfnamefont {M.}~\bibnamefont {Hashimoto}},
  \bibinfo {author} {\bibfnamefont {H.}~\bibnamefont {Karapetyan}}, \bibinfo
  {author} {\bibfnamefont {J.~D.}\ \bibnamefont {Koralek}}, \bibinfo {author}
  {\bibfnamefont {J.~P.}\ \bibnamefont {Hinton}}, \bibinfo {author}
  {\bibfnamefont {J.~P.}\ \bibnamefont {Testaud}}, \bibinfo {author}
  {\bibfnamefont {V.}~\bibnamefont {Nathan}}, \bibinfo {author} {\bibfnamefont
  {Y.}~\bibnamefont {Yoshida}}, \bibinfo {author} {\bibfnamefont
  {H.}~\bibnamefont {Yao}}, \bibinfo {author} {\bibfnamefont {K.}~\bibnamefont
  {Tanaka}}, \bibinfo {author} {\bibfnamefont {W.}~\bibnamefont {Meevasana}},
  \bibinfo {author} {\bibfnamefont {R.~G.}\ \bibnamefont {Moore}}, \bibinfo
  {author} {\bibfnamefont {D.~H.}\ \bibnamefont {Lu}}, \bibinfo {author}
  {\bibfnamefont {S.-K.}\ \bibnamefont {Mo}}, \bibinfo {author} {\bibfnamefont
  {M.}~\bibnamefont {Ishikado}}, \bibinfo {author} {\bibfnamefont
  {H.}~\bibnamefont {Eisaki}}, \bibinfo {author} {\bibfnamefont
  {Z.}~\bibnamefont {Hussain}}, \bibinfo {author} {\bibfnamefont {T.~P.}\
  \bibnamefont {Devereaux}}, \bibinfo {author} {\bibfnamefont {S.~A.}\
  \bibnamefont {Kivelson}}, \bibinfo {author} {\bibfnamefont {J.}~\bibnamefont
  {Orenstein}}, \bibinfo {author} {\bibfnamefont {A.}~\bibnamefont
  {Kapitulnik}}, \ and\ \bibinfo {author} {\bibfnamefont {Z.-X.}\ \bibnamefont
  {Shen}},\ }\href {\doibase 10.1126/science.1198415} {\bibfield  {journal}
  {\bibinfo  {journal} {Science}\ }\textbf {\bibinfo {volume} {331}},\ \bibinfo
  {pages} {1579} (\bibinfo {year} {2011})}\BibitemShut {NoStop}%
\bibitem [{\citenamefont {Kancharla}\ \emph {et~al.}(2008)\citenamefont
  {Kancharla}, \citenamefont {Kyung}, \citenamefont {Sénéchal}, \citenamefont
  {Civelli}, \citenamefont {Capone}, \citenamefont {Kotliar},\ and\
  \citenamefont {Tremblay}}]{kancharla_anomalous_2008}%
  \BibitemOpen
  \bibfield  {author} {\bibinfo {author} {\bibfnamefont {S.}~\bibnamefont
  {Kancharla}}, \bibinfo {author} {\bibfnamefont {B.}~\bibnamefont {Kyung}},
  \bibinfo {author} {\bibfnamefont {D.}~\bibnamefont {Sénéchal}}, \bibinfo
  {author} {\bibfnamefont {M.}~\bibnamefont {Civelli}}, \bibinfo {author}
  {\bibfnamefont {M.}~\bibnamefont {Capone}}, \bibinfo {author} {\bibfnamefont
  {G.}~\bibnamefont {Kotliar}}, \ and\ \bibinfo {author} {\bibfnamefont
  {A.-M.}\ \bibnamefont {Tremblay}},\ }\href {\doibase
  10.1103/PhysRevB.77.184516} {\bibfield  {journal} {\bibinfo  {journal}
  {Physical Review B}\ }\textbf {\bibinfo {volume} {77}},\ \bibinfo {pages}
  {184516} (\bibinfo {year} {2008})}\BibitemShut {NoStop}%
\bibitem [{\citenamefont {Mattheiss}(1990)}]{mattheiss_electronic_1990}%
  \BibitemOpen
  \bibfield  {author} {\bibinfo {author} {\bibfnamefont {L.~F.}\ \bibnamefont
  {Mattheiss}},\ }\href {\doibase 10.1103/PhysRevB.42.354} {\bibfield
  {journal} {\bibinfo  {journal} {Physical Review B}\ }\textbf {\bibinfo
  {volume} {42}},\ \bibinfo {pages} {354} (\bibinfo {year} {1990})}\BibitemShut
  {NoStop}%
\bibitem [{\citenamefont {Schabel}\ \emph {et~al.}(1998)\citenamefont
  {Schabel}, \citenamefont {Park}, \citenamefont {Matsuura}, \citenamefont
  {Shen}, \citenamefont {Bonn}, \citenamefont {Liang},\ and\ \citenamefont
  {Hardy}}]{schabel_angle-resolved_1998}%
  \BibitemOpen
  \bibfield  {author} {\bibinfo {author} {\bibfnamefont {M.~C.}\ \bibnamefont
  {Schabel}}, \bibinfo {author} {\bibfnamefont {C.-H.}\ \bibnamefont {Park}},
  \bibinfo {author} {\bibfnamefont {A.}~\bibnamefont {Matsuura}}, \bibinfo
  {author} {\bibfnamefont {Z.-X.}\ \bibnamefont {Shen}}, \bibinfo {author}
  {\bibfnamefont {D.~A.}\ \bibnamefont {Bonn}}, \bibinfo {author}
  {\bibfnamefont {R.}~\bibnamefont {Liang}}, \ and\ \bibinfo {author}
  {\bibfnamefont {W.~N.}\ \bibnamefont {Hardy}},\ }\href {\doibase
  10.1103/PhysRevB.57.6090} {\bibfield  {journal} {\bibinfo  {journal}
  {Physical Review B}\ }\textbf {\bibinfo {volume} {57}},\ \bibinfo {pages}
  {6090} (\bibinfo {year} {1998})}\BibitemShut {NoStop}%
\bibitem [{\citenamefont {Hoogenboom}\ \emph {et~al.}(2003)\citenamefont
  {Hoogenboom}, \citenamefont {Berthod}, \citenamefont {Peter}, \citenamefont
  {Fischer},\ and\ \citenamefont {Kordyuk}}]{hoogenboom_modeling_2003}%
  \BibitemOpen
  \bibfield  {author} {\bibinfo {author} {\bibfnamefont {B.~W.}\ \bibnamefont
  {Hoogenboom}}, \bibinfo {author} {\bibfnamefont {C.}~\bibnamefont {Berthod}},
  \bibinfo {author} {\bibfnamefont {M.}~\bibnamefont {Peter}}, \bibinfo
  {author} {\bibfnamefont {O.~.}\ \bibnamefont {Fischer}}, \ and\ \bibinfo
  {author} {\bibfnamefont {A.~A.}\ \bibnamefont {Kordyuk}},\ }\href {\doibase
  10.1103/PhysRevB.67.224502} {\bibfield  {journal} {\bibinfo  {journal}
  {Physical Review B}\ }\textbf {\bibinfo {volume} {67}},\ \bibinfo {pages}
  {224502} (\bibinfo {year} {2003})}\BibitemShut {NoStop}%
\bibitem [{\citenamefont {Comin}\ \emph {et~al.}(2015)\citenamefont {Comin},
  \citenamefont {Sutarto}, \citenamefont {Neto}, \citenamefont {Chauviere},
  \citenamefont {Liang}, \citenamefont {Hardy}, \citenamefont {Bonn},
  \citenamefont {He}, \citenamefont {Sawatzky},\ and\ \citenamefont
  {Damascelli}}]{comin_broken_2015}%
  \BibitemOpen
  \bibfield  {author} {\bibinfo {author} {\bibfnamefont {R.}~\bibnamefont
  {Comin}}, \bibinfo {author} {\bibfnamefont {R.}~\bibnamefont {Sutarto}},
  \bibinfo {author} {\bibfnamefont {E.~H. d.~S.}\ \bibnamefont {Neto}},
  \bibinfo {author} {\bibfnamefont {L.}~\bibnamefont {Chauviere}}, \bibinfo
  {author} {\bibfnamefont {R.}~\bibnamefont {Liang}}, \bibinfo {author}
  {\bibfnamefont {W.~N.}\ \bibnamefont {Hardy}}, \bibinfo {author}
  {\bibfnamefont {D.~A.}\ \bibnamefont {Bonn}}, \bibinfo {author}
  {\bibfnamefont {F.}~\bibnamefont {He}}, \bibinfo {author} {\bibfnamefont
  {G.~A.}\ \bibnamefont {Sawatzky}}, \ and\ \bibinfo {author} {\bibfnamefont
  {A.}~\bibnamefont {Damascelli}},\ }\href {\doibase 10.1126/science.1258399}
  {\bibfield  {journal} {\bibinfo  {journal} {Science}\ }\textbf {\bibinfo
  {volume} {347}},\ \bibinfo {pages} {1335} (\bibinfo {year}
  {2015})}\BibitemShut {NoStop}%
\bibitem [{\citenamefont {Mesaros}\ \emph {et~al.}(2016)\citenamefont
  {Mesaros}, \citenamefont {Fujita}, \citenamefont {Edkins}, \citenamefont
  {Hamidian}, \citenamefont {Eisaki}, \citenamefont {Uchida}, \citenamefont
  {Davis}, \citenamefont {Lawler},\ and\ \citenamefont
  {Kim}}]{mesaros_commensurate_2016}%
  \BibitemOpen
  \bibfield  {author} {\bibinfo {author} {\bibfnamefont {A.}~\bibnamefont
  {Mesaros}}, \bibinfo {author} {\bibfnamefont {K.}~\bibnamefont {Fujita}},
  \bibinfo {author} {\bibfnamefont {S.~D.}\ \bibnamefont {Edkins}}, \bibinfo
  {author} {\bibfnamefont {M.~H.}\ \bibnamefont {Hamidian}}, \bibinfo {author}
  {\bibfnamefont {H.}~\bibnamefont {Eisaki}}, \bibinfo {author} {\bibfnamefont
  {S.}~\bibnamefont {Uchida}}, \bibinfo {author} {\bibfnamefont {J.~C.~S.}\
  \bibnamefont {Davis}}, \bibinfo {author} {\bibfnamefont {M.~J.}\ \bibnamefont
  {Lawler}}, \ and\ \bibinfo {author} {\bibfnamefont {E.-A.}\ \bibnamefont
  {Kim}},\ }\href {http://arxiv.org/abs/1608.06180} {\bibfield  {journal}
  {\bibinfo  {journal} {arXiv:1608.06180 [cond-mat]}\ } (\bibinfo {year}
  {2016})},\ \bibinfo {note} {arXiv: 1608.06180}\BibitemShut {NoStop}%
\bibitem [{Note1()}]{Note1}%
  \BibitemOpen
  \bibinfo {note} {Non-nesting DW scenario can be justified given a
  sufficiently complicated interaction function $U(\protect \bm {\protect
  \mathrm {q}})$, because then the largest susceptibility $\chi _0 /
  (1-U(\protect \bm {\protect \mathrm {q}})\chi _0(\protect \bm {\protect
  \mathrm {q}}))$ is not necessarily at a nesting wave vector of the Fermi
  surface.}\BibitemShut {Stop}%
\bibitem [{\citenamefont {Chang}\ \emph {et~al.}(2012)\citenamefont {Chang},
  \citenamefont {Blackburn}, \citenamefont {Holmes}, \citenamefont
  {Christensen}, \citenamefont {Larsen}, \citenamefont {Mesot}, \citenamefont
  {Liang}, \citenamefont {Bonn}, \citenamefont {Hardy}, \citenamefont
  {Watenphul}, \citenamefont {Zimmermann}, \citenamefont {Forgan},\ and\
  \citenamefont {Hayden}}]{chang_direct_2012}%
  \BibitemOpen
  \bibfield  {author} {\bibinfo {author} {\bibfnamefont {J.}~\bibnamefont
  {Chang}}, \bibinfo {author} {\bibfnamefont {E.}~\bibnamefont {Blackburn}},
  \bibinfo {author} {\bibfnamefont {A.~T.}\ \bibnamefont {Holmes}}, \bibinfo
  {author} {\bibfnamefont {N.~B.}\ \bibnamefont {Christensen}}, \bibinfo
  {author} {\bibfnamefont {J.}~\bibnamefont {Larsen}}, \bibinfo {author}
  {\bibfnamefont {J.}~\bibnamefont {Mesot}}, \bibinfo {author} {\bibfnamefont
  {R.}~\bibnamefont {Liang}}, \bibinfo {author} {\bibfnamefont {D.~A.}\
  \bibnamefont {Bonn}}, \bibinfo {author} {\bibfnamefont {W.~N.}\ \bibnamefont
  {Hardy}}, \bibinfo {author} {\bibfnamefont {A.}~\bibnamefont {Watenphul}},
  \bibinfo {author} {\bibfnamefont {M.~v.}\ \bibnamefont {Zimmermann}},
  \bibinfo {author} {\bibfnamefont {E.~M.}\ \bibnamefont {Forgan}}, \ and\
  \bibinfo {author} {\bibfnamefont {S.~M.}\ \bibnamefont {Hayden}},\ }\href
  {\doibase 10.1038/nphys2456} {\bibfield  {journal} {\bibinfo  {journal}
  {Nature Physics}\ }\textbf {\bibinfo {volume} {8}},\ \bibinfo {pages} {871}
  (\bibinfo {year} {2012})}\BibitemShut {NoStop}%
\bibitem [{\citenamefont {Matsuba}\ \emph {et~al.}(2007)\citenamefont
  {Matsuba}, \citenamefont {Yoshizawa}, \citenamefont {Mochizuki},
  \citenamefont {Mochiku}, \citenamefont {Hirata},\ and\ \citenamefont
  {Nishida}}]{matsuba_anti-phase_2007}%
  \BibitemOpen
  \bibfield  {author} {\bibinfo {author} {\bibfnamefont {K.}~\bibnamefont
  {Matsuba}}, \bibinfo {author} {\bibfnamefont {S.}~\bibnamefont {Yoshizawa}},
  \bibinfo {author} {\bibfnamefont {Y.}~\bibnamefont {Mochizuki}}, \bibinfo
  {author} {\bibfnamefont {T.}~\bibnamefont {Mochiku}}, \bibinfo {author}
  {\bibfnamefont {K.}~\bibnamefont {Hirata}}, \ and\ \bibinfo {author}
  {\bibfnamefont {N.}~\bibnamefont {Nishida}},\ }\href {\doibase
  10.1143/JPSJ.76.063704} {\bibfield  {journal} {\bibinfo  {journal} {Journal
  of the Physical Society of Japan}\ }\textbf {\bibinfo {volume} {76}},\
  \bibinfo {pages} {063704} (\bibinfo {year} {2007})}\BibitemShut {NoStop}%
\bibitem [{\citenamefont {Kloss}\ \emph {et~al.}(2015)\citenamefont {Kloss},
  \citenamefont {Montiel},\ and\ \citenamefont {Pépin}}]{kloss_su2_2015}%
  \BibitemOpen
  \bibfield  {author} {\bibinfo {author} {\bibfnamefont {T.}~\bibnamefont
  {Kloss}}, \bibinfo {author} {\bibfnamefont {X.}~\bibnamefont {Montiel}}, \
  and\ \bibinfo {author} {\bibfnamefont {C.}~\bibnamefont {Pépin}},\ }\href
  {\doibase 10.1103/PhysRevB.91.205124} {\bibfield  {journal} {\bibinfo
  {journal} {Physical Review B}\ }\textbf {\bibinfo {volume} {91}},\ \bibinfo
  {pages} {205124} (\bibinfo {year} {2015})}\BibitemShut {NoStop}%
\bibitem [{\citenamefont {Montiel}\ \emph {et~al.}(2016)\citenamefont
  {Montiel}, \citenamefont {Kloss},\ and\ \citenamefont
  {Pépin}}]{montiel_angle_2016}%
  \BibitemOpen
  \bibfield  {author} {\bibinfo {author} {\bibfnamefont {X.}~\bibnamefont
  {Montiel}}, \bibinfo {author} {\bibfnamefont {T.}~\bibnamefont {Kloss}}, \
  and\ \bibinfo {author} {\bibfnamefont {C.}~\bibnamefont {Pépin}},\ }\href
  {http://arxiv.org/abs/1603.09178} {\bibfield  {journal} {\bibinfo  {journal}
  {arXiv:1603.09178 [cond-mat]}\ } (\bibinfo {year} {2016})},\ \bibinfo {note}
  {arXiv: 1603.09178}\BibitemShut {NoStop}%
\bibitem [{\citenamefont {Kreisel}\ \emph {et~al.}(2015)\citenamefont
  {Kreisel}, \citenamefont {Choubey}, \citenamefont {Berlijn}, \citenamefont
  {Ku}, \citenamefont {Andersen},\ and\ \citenamefont
  {Hirschfeld}}]{kreisel_interpretation_2015}%
  \BibitemOpen
  \bibfield  {author} {\bibinfo {author} {\bibfnamefont {A.}~\bibnamefont
  {Kreisel}}, \bibinfo {author} {\bibfnamefont {P.}~\bibnamefont {Choubey}},
  \bibinfo {author} {\bibfnamefont {T.}~\bibnamefont {Berlijn}}, \bibinfo
  {author} {\bibfnamefont {W.}~\bibnamefont {Ku}}, \bibinfo {author}
  {\bibfnamefont {B.}~\bibnamefont {Andersen}}, \ and\ \bibinfo {author}
  {\bibfnamefont {P.}~\bibnamefont {Hirschfeld}},\ }\href {\doibase
  10.1103/PhysRevLett.114.217002} {\bibfield  {journal} {\bibinfo  {journal}
  {Physical Review Letters}\ }\textbf {\bibinfo {volume} {114}},\ \bibinfo
  {pages} {217002} (\bibinfo {year} {2015})}\BibitemShut {NoStop}%
\bibitem [{\citenamefont {Choubey}\ \emph {et~al.}(2016)\citenamefont
  {Choubey}, \citenamefont {Tu}, \citenamefont {Lee},\ and\ \citenamefont
  {Hirschfeld}}]{choubey_incommensurate_2016}%
  \BibitemOpen
  \bibfield  {author} {\bibinfo {author} {\bibfnamefont {P.}~\bibnamefont
  {Choubey}}, \bibinfo {author} {\bibfnamefont {W.-L.}\ \bibnamefont {Tu}},
  \bibinfo {author} {\bibfnamefont {T.-K.}\ \bibnamefont {Lee}}, \ and\
  \bibinfo {author} {\bibfnamefont {P.~J.}\ \bibnamefont {Hirschfeld}},\ }\href
  {http://arxiv.org/abs/1609.07067} {\bibfield  {journal} {\bibinfo  {journal}
  {arXiv:1609.07067 [cond-mat]}\ } (\bibinfo {year} {2016})},\ \bibinfo {note}
  {arXiv: 1609.07067}\BibitemShut {NoStop}%
\bibitem [{\citenamefont {Norman}\ \emph {et~al.}(1998)\citenamefont {Norman},
  \citenamefont {Randeria}, \citenamefont {Ding},\ and\ \citenamefont
  {Campuzano}}]{norman_phenomenology_1998}%
  \BibitemOpen
  \bibfield  {author} {\bibinfo {author} {\bibfnamefont {M.~R.}\ \bibnamefont
  {Norman}}, \bibinfo {author} {\bibfnamefont {M.}~\bibnamefont {Randeria}},
  \bibinfo {author} {\bibfnamefont {H.}~\bibnamefont {Ding}}, \ and\ \bibinfo
  {author} {\bibfnamefont {J.~C.}\ \bibnamefont {Campuzano}},\ }\href {\doibase
  10.1103/PhysRevB.57.R11093} {\bibfield  {journal} {\bibinfo  {journal}
  {Physical Review B}\ }\textbf {\bibinfo {volume} {57}},\ \bibinfo {pages}
  {R11093} (\bibinfo {year} {1998})}\BibitemShut {NoStop}%
\bibitem [{\citenamefont {Kondo}\ \emph {et~al.}(2015)\citenamefont {Kondo},
  \citenamefont {Malaeb}, \citenamefont {Ishida}, \citenamefont {Sasagawa},
  \citenamefont {Sakamoto}, \citenamefont {Takeuchi}, \citenamefont {Tohyama},\
  and\ \citenamefont {Shin}}]{kondo_point_2015}%
  \BibitemOpen
  \bibfield  {author} {\bibinfo {author} {\bibfnamefont {T.}~\bibnamefont
  {Kondo}}, \bibinfo {author} {\bibfnamefont {W.}~\bibnamefont {Malaeb}},
  \bibinfo {author} {\bibfnamefont {Y.}~\bibnamefont {Ishida}}, \bibinfo
  {author} {\bibfnamefont {T.}~\bibnamefont {Sasagawa}}, \bibinfo {author}
  {\bibfnamefont {H.}~\bibnamefont {Sakamoto}}, \bibinfo {author}
  {\bibfnamefont {T.}~\bibnamefont {Takeuchi}}, \bibinfo {author}
  {\bibfnamefont {T.}~\bibnamefont {Tohyama}}, \ and\ \bibinfo {author}
  {\bibfnamefont {S.}~\bibnamefont {Shin}},\ }\href {\doibase
  10.1038/ncomms8699} {\bibfield  {journal} {\bibinfo  {journal} {Nature
  Communications}\ }\textbf {\bibinfo {volume} {6}} (\bibinfo {year} {2015}),\
  10.1038/ncomms8699}\BibitemShut {NoStop}%
\bibitem [{\citenamefont {LeBlanc}(2014)}]{leblanc_signatures_2014}%
  \BibitemOpen
  \bibfield  {author} {\bibinfo {author} {\bibfnamefont {J.~P.~F.}\
  \bibnamefont {LeBlanc}},\ }\href {\doibase 10.1088/1367-2630/16/11/113034}
  {\bibfield  {journal} {\bibinfo  {journal} {New Journal of Physics}\ }\textbf
  {\bibinfo {volume} {16}},\ \bibinfo {pages} {113034} (\bibinfo {year}
  {2014})}\BibitemShut {NoStop}%
\bibitem [{\citenamefont {Borne}\ \emph {et~al.}(2010)\citenamefont {Borne},
  \citenamefont {Carbotte},\ and\ \citenamefont
  {Nicol}}]{borne_signature_2010}%
  \BibitemOpen
  \bibfield  {author} {\bibinfo {author} {\bibfnamefont {A.~J.~H.}\
  \bibnamefont {Borne}}, \bibinfo {author} {\bibfnamefont {J.~P.}\ \bibnamefont
  {Carbotte}}, \ and\ \bibinfo {author} {\bibfnamefont {E.~J.}\ \bibnamefont
  {Nicol}},\ }\href {\doibase 10.1103/PhysRevB.82.024521} {\bibfield  {journal}
  {\bibinfo  {journal} {Physical Review B}\ }\textbf {\bibinfo {volume} {82}},\
  \bibinfo {pages} {024521} (\bibinfo {year} {2010})}\BibitemShut {NoStop}%
\bibitem [{\citenamefont {Pushp}\ \emph {et~al.}(2009)\citenamefont {Pushp},
  \citenamefont {Parker}, \citenamefont {Pasupathy}, \citenamefont {Gomes},
  \citenamefont {Ono}, \citenamefont {Wen}, \citenamefont {Xu}, \citenamefont
  {Gu},\ and\ \citenamefont {Yazdani}}]{pushp_extending_2009}%
  \BibitemOpen
  \bibfield  {author} {\bibinfo {author} {\bibfnamefont {A.}~\bibnamefont
  {Pushp}}, \bibinfo {author} {\bibfnamefont {C.~V.}\ \bibnamefont {Parker}},
  \bibinfo {author} {\bibfnamefont {A.~N.}\ \bibnamefont {Pasupathy}}, \bibinfo
  {author} {\bibfnamefont {K.~K.}\ \bibnamefont {Gomes}}, \bibinfo {author}
  {\bibfnamefont {S.}~\bibnamefont {Ono}}, \bibinfo {author} {\bibfnamefont
  {J.}~\bibnamefont {Wen}}, \bibinfo {author} {\bibfnamefont {Z.}~\bibnamefont
  {Xu}}, \bibinfo {author} {\bibfnamefont {G.}~\bibnamefont {Gu}}, \ and\
  \bibinfo {author} {\bibfnamefont {A.}~\bibnamefont {Yazdani}},\ }\href
  {\doibase 10.1126/science.1174338} {\bibfield  {journal} {\bibinfo  {journal}
  {Science}\ }\textbf {\bibinfo {volume} {324}},\ \bibinfo {pages} {1689}
  (\bibinfo {year} {2009})}\BibitemShut {NoStop}%
\bibitem [{Note2()}]{Note2}%
  \BibitemOpen
  \bibinfo {note} {The corresponding doping (approximately $p=0.2$) is not
  really relevant; we could tune the band structure, the AF mean-field and/or
  the DWs wave vector (here $\protect \frac {q}{L}=\protect \frac {1}{3}$), to
  fit the experiments, as in Ref.~\protect \rev@citealpnum
  {atkinson_charge_2015}.}\BibitemShut {Stop}%
\bibitem [{Note3()}]{Note3}%
  \BibitemOpen
  \bibinfo {note} {This is due to the nature of the $s'$PDW, which couples the
  band energy $\xi (\protect \bm {\protect \mathrm {k}})$ with $-\xi (\protect
  \bm {\protect \mathrm {k}}+\protect \bm {\protect \mathrm
  {Q}})$.}\BibitemShut {Stop}%
\bibitem [{\citenamefont {Kordyuk}(2015)}]{kordyuk_pseudogap_2015}%
  \BibitemOpen
  \bibfield  {author} {\bibinfo {author} {\bibfnamefont {A.~A.}\ \bibnamefont
  {Kordyuk}},\ }\href {\doibase 10.1063/1.4919371} {\bibfield  {journal}
  {\bibinfo  {journal} {Low Temperature Physics}\ }\textbf {\bibinfo {volume}
  {41}},\ \bibinfo {pages} {319} (\bibinfo {year} {2015})}\BibitemShut
  {NoStop}%
\bibitem [{\citenamefont {Hayward}\ \emph {et~al.}(2014)\citenamefont
  {Hayward}, \citenamefont {Hawthorn}, \citenamefont {Melko},\ and\
  \citenamefont {Sachdev}}]{hayward_angular_2014}%
  \BibitemOpen
  \bibfield  {author} {\bibinfo {author} {\bibfnamefont {L.~E.}\ \bibnamefont
  {Hayward}}, \bibinfo {author} {\bibfnamefont {D.~G.}\ \bibnamefont
  {Hawthorn}}, \bibinfo {author} {\bibfnamefont {R.~G.}\ \bibnamefont {Melko}},
  \ and\ \bibinfo {author} {\bibfnamefont {S.}~\bibnamefont {Sachdev}},\ }\href
  {\doibase 10.1126/science.1246310} {\bibfield  {journal} {\bibinfo  {journal}
  {Science}\ }\textbf {\bibinfo {volume} {343}},\ \bibinfo {pages} {1336}
  (\bibinfo {year} {2014})}\BibitemShut {NoStop}%
\bibitem [{\citenamefont {Fratino}\ \emph {et~al.}(2016)\citenamefont
  {Fratino}, \citenamefont {Sémon}, \citenamefont {Sordi},\ and\ \citenamefont
  {Tremblay}}]{fratino_organizing_2016}%
  \BibitemOpen
  \bibfield  {author} {\bibinfo {author} {\bibfnamefont {L.}~\bibnamefont
  {Fratino}}, \bibinfo {author} {\bibfnamefont {P.}~\bibnamefont {Sémon}},
  \bibinfo {author} {\bibfnamefont {G.}~\bibnamefont {Sordi}}, \ and\ \bibinfo
  {author} {\bibfnamefont {A.-M.~S.}\ \bibnamefont {Tremblay}},\ }\href
  {\doibase 10.1038/srep22715} {\bibfield  {journal} {\bibinfo  {journal}
  {Scientific Reports}\ }\textbf {\bibinfo {volume} {6}},\ \bibinfo {pages}
  {22715} (\bibinfo {year} {2016})}\BibitemShut {NoStop}%
\bibitem [{\citenamefont {Tu}\ and\ \citenamefont
  {Lee}(2016)}]{tu_genesis_2016}%
  \BibitemOpen
  \bibfield  {author} {\bibinfo {author} {\bibfnamefont {W.-L.}\ \bibnamefont
  {Tu}}\ and\ \bibinfo {author} {\bibfnamefont {T.-K.}\ \bibnamefont {Lee}},\
  }\href {\doibase 10.1038/srep18675} {\bibfield  {journal} {\bibinfo
  {journal} {Scientific Reports}\ }\textbf {\bibinfo {volume} {6}},\ \bibinfo
  {pages} {18675} (\bibinfo {year} {2016})}\BibitemShut {NoStop}%
\end{thebibliography}

\end{document}